\documentclass[manuscripts]{aastex631}

\usepackage{graphicx}

\usepackage{url}
\usepackage{graphicx}
\usepackage{float}
\usepackage{amssymb}
\usepackage{upgreek}
\usepackage{enumerate}
\usepackage{txfonts}
\usepackage{graphicx}
\usepackage{subfigure}
\usepackage{multirow}
\usepackage[T1]{fontenc}
\let\tablenum\relax
\usepackage{siunitx}
\usepackage{color}
\usepackage{times}
\usepackage{blindtext}

\newcommand{\insight}{\textit{Insight}-HXMT}
\newcommand{\sgr}{\mbox{SGR~J1935+2154~}}
\newcommand{\sgrnos}{\mbox{SGR~J1935+2154}}

\begin{document}


\title{Quasi-periodic oscillations of the X-ray burst from the magnetar SGR~J1935+2154 and associated with the fast radio burst FRB~200428}

\correspondingauthor{Shuang-Nan Zhang}
\email{zhangsn@ihep.ac.cn}

\author[0000-0003-4585-589X]{Xiaobo Li}
\affiliation{Key Laboratory for Particle Astrophysics, Institute of High Energy Physics, Chinese Academy of Sciences, 19B Yuquan Road, Beijing 100049, China}

\author[0000-0002-3776-4536]{Mingyu Ge}
\affiliation{Key Laboratory for Particle Astrophysics, Institute of High Energy Physics, Chinese Academy of Sciences, 19B Yuquan Road, Beijing 100049, China}

\author{Lin Lin}
\affiliation{Department of Astronomy, Beijing Normal University, Beijing 100088, China}

\author[0000-0001-5586-1017]{Shuang-Nan Zhang\textsuperscript{*}}
\affiliation{Key Laboratory for Particle Astrophysics, Institute of High Energy Physics, Chinese Academy of Sciences, 19B Yuquan Road, Beijing 100049, China}
\affiliation{University of Chinese Academy of Sciences, Chinese Academy of Sciences, Beijing 100049, China}

\author[0000-0003-0274-3396]{Liming Song}
\affiliation{Key Laboratory for Particle Astrophysics, Institute of High Energy Physics, Chinese Academy of Sciences, 19B Yuquan Road, Beijing 100049, China}
\affiliation{University of Chinese Academy of Sciences, Chinese Academy of Sciences, Beijing 100049, China}

\author{Xuelei Cao}
\affiliation{Key Laboratory for Particle Astrophysics, Institute of High Energy Physics, Chinese Academy of Sciences, 19B Yuquan Road, Beijing 100049, China}

\author{Bing Zhang}
\affiliation{Department of Physics and Astronomy, University of Nevad, Las Vegas, NV 89154, USA}

\author[0000-0003-3248-6087]{Fangjun Lu}
\affiliation{Key Laboratory for Particle Astrophysics, Institute of High Energy Physics, Chinese Academy of Sciences, 19B Yuquan Road, Beijing 100049, China}
\affil{Key Laboratory of Stellar and Interstellar Physics and School of Physics and Optoelectronics, Xiangtan University, Xiangtan 411105, Hunan, China}

\author[0000-0002-8476-9217]{Yupeng Xu}
\affiliation{Key Laboratory for Particle Astrophysics, Institute of High Energy Physics, Chinese Academy of Sciences, 19B Yuquan Road, Beijing 100049, China}
\affiliation{University of Chinese Academy of Sciences, Chinese Academy of Sciences, Beijing 100049, China}

\author{Shaolin Xiong}
\affiliation{Key Laboratory for Particle Astrophysics, Institute of High Energy Physics, Chinese Academy of Sciences, 19B Yuquan Road, Beijing 100049, China}
\affiliation{University of Chinese Academy of Sciences, Chinese Academy of Sciences, Beijing 100049, China}

\author[0000-0003-3127-0110]{Youli Tuo}
\affiliation{Key Laboratory for Particle Astrophysics, Institute of High Energy Physics, Chinese Academy of Sciences, 19B Yuquan Road, Beijing 100049, China}

\author{Ying Tan}
\affiliation{Key Laboratory for Particle Astrophysics, Institute of High Energy Physics, Chinese Academy of Sciences, 19B Yuquan Road, Beijing 100049, China}

\author{Weichun Jiang}
\affiliation{Key Laboratory for Particle Astrophysics, Institute of High Energy Physics, Chinese Academy of Sciences, 19B Yuquan Road, Beijing 100049, China}

\author{Jinlu Qu}
\affiliation{Key Laboratory for Particle Astrophysics, Institute of High Energy Physics, Chinese Academy of Sciences, 19B Yuquan Road, Beijing 100049, China}

\author{Shu Zhang}
\affiliation{Key Laboratory for Particle Astrophysics, Institute of High Energy Physics, Chinese Academy of Sciences, 19B Yuquan Road, Beijing 100049, China}

\author{Lingjun Wang}
\affiliation{Key Laboratory for Particle Astrophysics, Institute of High Energy Physics, Chinese Academy of Sciences, 19B Yuquan Road, Beijing 100049, China}

\author{Jieshuang Wang}
\affiliation{Max-Planck-Institut f\"ur Kernphysik, Saupfercheckweg 1, D-69117 Heidelberg, Germany}

\author{Binbin Zhang}
\affiliation{School of Astronomy and Space Science, Nanjing University, Nanjing 210023, China}

\author{Peng Zhang}
\affil{Key Laboratory for Particle Astrophysics, Institute of High Energy Physics, Chinese Academy of Sciences, 19B Yuquan Road, Beijing 100049, China}

\author{Chengkui Li}
\affil{Key Laboratory for Particle Astrophysics, Institute of High Energy Physics, Chinese Academy of Sciences, 19B Yuquan Road, Beijing 100049, China}

\author[0000-0002-4834-9637]{Congzhan Liu}
\affiliation{Key Laboratory for Particle Astrophysics, Institute of High Energy Physics, Chinese Academy of Sciences, 19B Yuquan Road, Beijing 100049, China}

\author{Tipei Li}
\affil{Key Laboratory for Particle Astrophysics, Institute of High Energy Physics, Chinese Academy of Sciences, 19B Yuquan Road, Beijing 100049, China}
\affil{Department of Astronomy, Tsinghua University, Beijing 100084, China}
\affil{University of Chinese Academy of Sciences, Chinese Academy of Sciences, Beijing 100049, China}

\author{Qingcui Bu}
\affiliation{Key Laboratory for Particle Astrophysics, Institute of High Energy Physics, Chinese Academy of Sciences, 19B Yuquan Road, Beijing 100049, China}

\author{Ce Cai}
\affiliation{Key Laboratory for Particle Astrophysics, Institute of High Energy Physics, Chinese Academy of Sciences, 19B Yuquan Road, Beijing 100049, China}

\author{Yong Chen}
\affiliation{Key Laboratory for Particle Astrophysics, Institute of High Energy Physics, Chinese Academy of Sciences, 19B Yuquan Road, Beijing 100049, China}

\author{Yupeng Chen}
\affil{Key Laboratory for Particle Astrophysics, Institute of High Energy Physics, Chinese Academy of Sciences, 19B Yuquan Road, Beijing 100049, China}

\author{Zhi Chang}
\affil{Key Laboratory for Particle Astrophysics, Institute of High Energy Physics, Chinese Academy of Sciences, 19B Yuquan Road, Beijing 100049, China}

\author{Li Chen}
\affil{Department of Astronomy, Beijing Normal University, Beijing 100088, China}

\author{Tianxian Chen}
\affil{Key Laboratory for Particle Astrophysics, Institute of High Energy Physics, Chinese Academy of Sciences, 19B Yuquan Road, Beijing 100049, China}

\author{Yibao Chen}
\affil{Department of Physics, Tsinghua University, Beijing 100084, China}

\author{Weiwei Cui}
\affil{Key Laboratory for Particle Astrophysics, Institute of High Energy Physics, Chinese Academy of Sciences, 19B Yuquan Road, Beijing 100049, China}

\author{Yuanyuan Du}
\affil{Key Laboratory for Particle Astrophysics, Institute of High Energy Physics, Chinese Academy of Sciences, 19B Yuquan Road, Beijing 100049, China}

\author{Guanhua Gao}
\affil{Key Laboratory for Particle Astrophysics, Institute of High Energy Physics, Chinese Academy of Sciences, 19B Yuquan Road, Beijing 100049, China}
\affil{University of Chinese Academy of Sciences, Chinese Academy of Sciences, Beijing 100049, China}

\author{He Gao}
\affil{Key Laboratory for Particle Astrophysics, Institute of High Energy Physics, Chinese Academy of Sciences, 19B Yuquan Road, Beijing 100049, China}
\affil{University of Chinese Academy of Sciences, Chinese Academy of Sciences, Beijing 100049, China}

\author{Yudong Gu}
\affil{Key Laboratory for Particle Astrophysics, Institute of High Energy Physics, Chinese Academy of Sciences, 19B Yuquan Road, Beijing 100049, China}

\author{Ju Guan}
\affil{Key Laboratory for Particle Astrophysics, Institute of High Energy Physics, Chinese Academy of Sciences, 19B Yuquan Road, Beijing 100049, China}

\author{Chengcheng Guo}
\affil{Key Laboratory for Particle Astrophysics, Institute of High Energy Physics, Chinese Academy of Sciences, 19B Yuquan Road, Beijing 100049, China}
\affil{University of Chinese Academy of Sciences, Chinese Academy of Sciences, Beijing 100049, China}

\author{Dawei Han}
\affil{Key Laboratory for Particle Astrophysics, Institute of High Energy Physics, Chinese Academy of Sciences, 19B Yuquan Road, Beijing 100049, China}

\author{Yue Huang}
\affil{Key Laboratory for Particle Astrophysics, Institute of High Energy Physics, Chinese Academy of Sciences, 19B Yuquan Road, Beijing 100049, China}

\author{Jia Huo}
\affil{Key Laboratory for Particle Astrophysics, Institute of High Energy Physics, Chinese Academy of Sciences, 19B Yuquan Road, Beijing 100049, China}

\author{Shumei Jia}
\affil{Key Laboratory for Particle Astrophysics, Institute of High Energy Physics, Chinese Academy of Sciences, 19B Yuquan Road, Beijing 100049, China}

\author{Jing Jin}
\affil{Key Laboratory for Particle Astrophysics, Institute of High Energy Physics, Chinese Academy of Sciences, 19B Yuquan Road, Beijing 100049, China}

\author{Lingda Kong}
\affil{Key Laboratory for Particle Astrophysics, Institute of High Energy Physics, Chinese Academy of Sciences, 19B Yuquan Road, Beijing 100049, China}
\affil{University of Chinese Academy of Sciences, Chinese Academy of Sciences, Beijing 100049, China}

\author{Bing Li}
\affil{Key Laboratory for Particle Astrophysics, Institute of High Energy Physics, Chinese Academy of Sciences, 19B Yuquan Road, Beijing 100049, China}

\author{Gang Li}
\affil{Key Laboratory for Particle Astrophysics, Institute of High Energy Physics, Chinese Academy of Sciences, 19B Yuquan Road, Beijing 100049, China}

\author{Wei Li}
\affil{Key Laboratory for Particle Astrophysics, Institute of High Energy Physics, Chinese Academy of Sciences, 19B Yuquan Road, Beijing 100049, China}

\author{Xian Li}
\affil{Key Laboratory for Particle Astrophysics, Institute of High Energy Physics, Chinese Academy of Sciences, 19B Yuquan Road, Beijing 100049, China}

\author{Xufang Li}
\affil{Key Laboratory for Particle Astrophysics, Institute of High Energy Physics, Chinese Academy of Sciences, 19B Yuquan Road, Beijing 100049, China}

\author{Zhengwei Li}
\affil{Key Laboratory for Particle Astrophysics, Institute of High Energy Physics, Chinese Academy of Sciences, 19B Yuquan Road, Beijing 100049, China}

\author{Xiaohua Liang}
\affil{Key Laboratory for Particle Astrophysics, Institute of High Energy Physics, Chinese Academy of Sciences, 19B Yuquan Road, Beijing 100049, China}

\author{Jinyuan Liao}
\affil{Key Laboratory for Particle Astrophysics, Institute of High Energy Physics, Chinese Academy of Sciences, 19B Yuquan Road, Beijing 100049, China}

\author{Hexin Liu}
\affil{Key Laboratory for Particle Astrophysics, Institute of High Energy Physics, Chinese Academy of Sciences, 19B Yuquan Road, Beijing 100049, China}
\affil{University of Chinese Academy of Sciences, Chinese Academy of Sciences, Beijing 100049, China}

\author{Hongwei Liu}
\affil{Key Laboratory for Particle Astrophysics, Institute of High Energy Physics, Chinese Academy of Sciences, 19B Yuquan Road, Beijing 100049, China}

\author{Xiaojing Liu}
\affil{Key Laboratory for Particle Astrophysics, Institute of High Energy Physics, Chinese Academy of Sciences, 19B Yuquan Road, Beijing 100049, China}

\author{Xuefeng Lu}
\affil{Key Laboratory for Particle Astrophysics, Institute of High Energy Physics, Chinese Academy of Sciences, 19B Yuquan Road, Beijing 100049, China}

\author{Qi Luo}
\affil{Key Laboratory for Particle Astrophysics, Institute of High Energy Physics, Chinese Academy of Sciences, 19B Yuquan Road, Beijing 100049, China}
\affil{University of Chinese Academy of Sciences, Chinese Academy of Sciences, Beijing 100049, China}

\author{Tao Luo}
\affil{Key Laboratory for Particle Astrophysics, Institute of High Energy Physics, Chinese Academy of Sciences, 19B Yuquan Road, Beijing 100049, China}

\author{Binyuan Ma}
\affil{Key Laboratory for Particle Astrophysics, Institute of High Energy Physics, Chinese Academy of Sciences, 19B Yuquan Road, Beijing 100049, China}
\affil{University of Chinese Academy of Sciences, Chinese Academy of Sciences, Beijing 100049, China}

\author{RuiCan Ma}
\affil{Key Laboratory for Particle Astrophysics, Institute of High Energy Physics, Chinese Academy of Sciences, 19B Yuquan Road, Beijing 100049, China}
\affil{University of Chinese Academy of Sciences, Chinese Academy of Sciences, Beijing 100049, China}

\author{Xiang Ma}
\affil{Key Laboratory for Particle Astrophysics, Institute of High Energy Physics, Chinese Academy of Sciences, 19B Yuquan Road, Beijing 100049, China}

\author{Bin Meng}
\affil{Key Laboratory for Particle Astrophysics, Institute of High Energy Physics, Chinese Academy of Sciences, 19B Yuquan Road, Beijing 100049, China}

\author{Yi Nang}
\affil{Key Laboratory for Particle Astrophysics, Institute of High Energy Physics, Chinese Academy of Sciences, 19B Yuquan Road, Beijing 100049, China}
\affil{University of Chinese Academy of Sciences, Chinese Academy of Sciences, Beijing 100049, China}

\author{Jianyin Nie}
\affil{Key Laboratory for Particle Astrophysics, Institute of High Energy Physics, Chinese Academy of Sciences, 19B Yuquan Road, Beijing 100049, China}

\author{Ge Ou}
\affil{Computing Division, Institute of High Energy Physics, Chinese Academy of Sciences, 19B Yuquan Road, Beijing 100049, China}

\author{Xiaoqin Ren}
\affil{Key Laboratory for Particle Astrophysics, Institute of High Energy Physics, Chinese Academy of Sciences, 19B Yuquan Road, Beijing 100049, China}
\affil{University of Chinese Academy of Sciences, Chinese Academy of Sciences, Beijing 100049, China}

\author{Na Sai}
\affil{Key Laboratory for Particle Astrophysics, Institute of High Energy Physics, Chinese Academy of Sciences, 19B Yuquan Road, Beijing 100049, China}
\affil{University of Chinese Academy of Sciences, Chinese Academy of Sciences, Beijing 100049, China}

\author{Xinying Song}
\affil{Key Laboratory for Particle Astrophysics, Institute of High Energy Physics, Chinese Academy of Sciences, 19B Yuquan Road, Beijing 100049, China}

\author{Liang Sun}
\affil{Key Laboratory for Particle Astrophysics, Institute of High Energy Physics, Chinese Academy of Sciences, 19B Yuquan Road, Beijing 100049, China}

\author{Lian Tao}
\affil{Key Laboratory for Particle Astrophysics, Institute of High Energy Physics, Chinese Academy of Sciences, 19B Yuquan Road, Beijing 100049, China}

\author{Chen Wang}
\affil{Key Laboratory of Space Astronomy and Technology, National Astronomical Observatories, Chinese
Academy of Sciences, Beijing 100012, China}
\affil{University of Chinese Academy of Sciences, Chinese Academy of Sciences, Beijing 100049, China}

\author{Pengju Wang}
\affil{Key Laboratory for Particle Astrophysics, Institute of High Energy Physics, Chinese Academy of Sciences, 19B Yuquan Road, Beijing 100049, China}
\affil{University of Chinese Academy of Sciences, Chinese Academy of Sciences, Beijing 100049, China}

\author{Wenshuai Wang}
\affil{Key Laboratory for Particle Astrophysics, Institute of High Energy Physics, Chinese Academy of Sciences, 19B Yuquan Road, Beijing 100049, China}

\author{Yusa Wang}
\affil{Key Laboratory for Particle Astrophysics, Institute of High Energy Physics, Chinese Academy of Sciences, 19B Yuquan Road, Beijing 100049, China}

\author{Xiangyang Wen}
\affil{Key Laboratory for Particle Astrophysics, Institute of High Energy Physics, Chinese Academy of Sciences, 19B Yuquan Road, Beijing 100049, China}

\author{Bobing Wu}
\affil{Key Laboratory for Particle Astrophysics, Institute of High Energy Physics, Chinese Academy of Sciences, 19B Yuquan Road, Beijing 100049, China}

\author{Baiyang Wu}
\affil{Key Laboratory for Particle Astrophysics, Institute of High Energy Physics, Chinese Academy of Sciences, 19B Yuquan Road, Beijing 100049, China}
\affil{University of Chinese Academy of Sciences, Chinese Academy of Sciences, Beijing 100049, China}

\author{Mei Wu}
\affil{Key Laboratory for Particle Astrophysics, Institute of High Energy Physics, Chinese Academy of Sciences, 19B Yuquan Road, Beijing 100049, China}

\author{Shuo Xiao}
\affil{Key Laboratory for Particle Astrophysics, Institute of High Energy Physics, Chinese Academy of Sciences, 19B Yuquan Road, Beijing 100049, China}
\affil{University of Chinese Academy of Sciences, Chinese Academy of Sciences, Beijing 100049, China}

\author{Sheng Yang}
\affil{Key Laboratory for Particle Astrophysics, Institute of High Energy Physics, Chinese Academy of Sciences, 19B Yuquan Road, Beijing 100049, China}

\author{Yanji Yang}
\affil{Key Laboratory for Particle Astrophysics, Institute of High Energy Physics, Chinese Academy of Sciences, 19B Yuquan Road, Beijing 100049, China}

\author{Qibin Yi}
\affil{Key Laboratory for Particle Astrophysics, Institute of High Energy Physics, Chinese Academy of Sciences, 19B Yuquan Road, Beijing 100049, China}
\affil{School of Physics and Optoelectronics, Xiangtan University, Yuhu District, Xiangtan, Hunan, 411105, China}

\author{Qianqing Yin}
\affil{Key Laboratory for Particle Astrophysics, Institute of High Energy Physics, Chinese Academy of Sciences, 19B Yuquan Road, Beijing 100049, China}

\author{Yuan You}
\affil{Key Laboratory for Particle Astrophysics, Institute of High Energy Physics, Chinese Academy of Sciences, 19B Yuquan Road, Beijing 100049, China}
\affil{University of Chinese Academy of Sciences, Chinese Academy of Sciences, Beijing 100049, China}

\author{Wei Yu}
\affil{Key Laboratory for Particle Astrophysics, Institute of High Energy Physics, Chinese Academy of Sciences, 19B Yuquan Road, Beijing 100049, China}
\affil{University of Chinese Academy of Sciences, Chinese Academy of Sciences, Beijing 100049, China}

\author{Fan Zhang}
\affil{Key Laboratory for Particle Astrophysics, Institute of High Energy Physics, Chinese Academy of Sciences, 19B Yuquan Road, Beijing 100049, China}

\author{Hongmei Zhang}
\affil{Key Laboratory for Particle Astrophysics, Institute of High Energy Physics, Chinese Academy of Sciences, 19B Yuquan Road, Beijing 100049, China}

\author{Juan Zhang}
\affil{Key Laboratory for Particle Astrophysics, Institute of High Energy Physics, Chinese Academy of Sciences, 19B Yuquan Road, Beijing 100049, China}

\author{Wanchang Zhang}
\affil{Key Laboratory for Particle Astrophysics, Institute of High Energy Physics, Chinese Academy of Sciences, 19B Yuquan Road, Beijing 100049, China}

\author{Wei Zhang}
\affil{Key Laboratory for Particle Astrophysics, Institute of High Energy Physics, Chinese Academy of Sciences, 19B Yuquan Road, Beijing 100049, China}
\affil{University of Chinese Academy of Sciences, Chinese Academy of Sciences, Beijing 100049, China}

\author{Yifei Zhang}
\affil{Key Laboratory for Particle Astrophysics, Institute of High Energy Physics, Chinese Academy of Sciences, 19B Yuquan Road, Beijing 100049, China}

\author{Yuanhang Zhang}
\affil{Key Laboratory for Particle Astrophysics, Institute of High Energy Physics, Chinese Academy of Sciences, 19B Yuquan Road, Beijing 100049, China}
\affil{University of Chinese Academy of Sciences, Chinese Academy of Sciences, Beijing 100049, China}

\author{Haisheng Zhao}
\affil{Key Laboratory for Particle Astrophysics, Institute of High Energy Physics, Chinese Academy of Sciences, 19B Yuquan Road, Beijing 100049, China}

\author{Xiaofan Zhao}
\affil{Key Laboratory for Particle Astrophysics, Institute of High Energy Physics, Chinese Academy of Sciences, 19B Yuquan Road, Beijing 100049, China}
\affil{University of Chinese Academy of Sciences, Chinese Academy of Sciences, Beijing 100049, China}

\author{Shijie Zheng}
\affil{Key Laboratory for Particle Astrophysics, Institute of High Energy Physics, Chinese Academy of Sciences, 19B Yuquan Road, Beijing 100049, China}

\author{Dengke Zhou}
\affil{Key Laboratory for Particle Astrophysics, Institute of High Energy Physics, Chinese Academy of Sciences, 19B Yuquan Road, Beijing 100049, China}
\affil{University of Chinese Academy of Sciences, Chinese Academy of Sciences, Beijing 100049, China}

\begin{abstract}
The origin(s) and mechanism(s) of fast radio bursts (FRBs), which are short radio pulses from cosmological distances, have remained a major puzzle since their discovery. We report a strong Quasi-Periodic Oscillation (QPO) of $\sim$40\,Hz in the X-ray burst from the magnetar \sgr{} and associated with FRB~200428, significantly detected with the Hard X-ray Modulation Telescope (\textit{Insight}-HXMT) and also hinted by the Konus-Wind data. QPOs from magnetar bursts have only been rarely detected; our 3.4\,$\sigma$ (p-value is 2.9e-4) detection of the QPO reported here reveals the strongest QPO signal observed from magnetars (except in some very rare giant flares), making this X-ray burst unique among magnetar bursts. The two X-ray spikes coinciding with the two FRB pulses are also among the peaks of the QPO. Our results suggest that at least some FRBs are related to strong oscillation processes of neutron stars.
We also show that we may overestimate the significance of the QPO signal and underestimate the errors of QPO parameters if QPO exists only in a fraction of the time series of a X-ray burst which we use to calculate the Leahy-normalized periodogram.
\end{abstract}


 \section{Introduction}
A major class of the sources of fast radio bursts (FRBs) \citep{2007Sci...318..777L} have long been proposed and recently been proved to be magnetars, which are neutron stars with surface magnetic field higher than $10^{14}$ Gauss \citep{ck98}. \sgr{} is a Galactic magnetar with the most frequent bursting activities \citep{israel2016,lin2020}. It went into burst active episodes in 2014, 2015, 2016, 2019, 2020, 2021, and 2022 \citep{younes2017,lin2020,2020ApJ...902L..43L,2021GCN.29377....1W,2022GCN.31443....1X}.
In the active episodes of 2020, a giant radio burst containing two pulses, which coincides in time with a bright X-ray burst, has been reported from it \citep{2020Natur.587...54C,2020Natur.587...59B,2020ApJ...898L..29M,2021NatAs...5..378L,2021NatAs...5..372R,2021NatAs...5..401T}. The radio burst, denoted as FRB~200428, is the first FRB detected in other wavelengths, and thus \sgr{} becomes the first counterpart of a FRB. This immediately establishes that at least some cosmological FRBs are produced during magnetar bursts. However, the exact mechanism behind this type of mysterious phenomena is unclear. It is well established that the majority of X-ray bursts from \sgr{} do not come with radio emission down to very low fluence \citep{2020Natur.587...63L,2021NatAs...5..414K} and weak radio bursts do not come with bright X-ray bursts \citep{fastatel2020b,2021NatAs...5..414K}. Therefore this peculiar X-ray burst and its association with FRB~200428 remains the only known FRB-magnetar connection. This X-ray burst is not a typical magnetar short burst, due to its non-thermal X-ray spectrum \citep{zhang2020,2021NatAs...5..378L,2021NatAs...5..408Y}. Therefore, normal mechanisms producing magnetar X-ray bursts will not generate (observable) FRBs, or at least not frequently.

Short X-ray bursts and flares from a magnetar can be generated from starquakes \citep{td95}. The damping of crustal oscillations due to core –crust coupling of the neutron star would leave imprints in the form of Quasi-periodic Oscillations (QPOs) in the temporal profiles of bursts \citep{huppenkothen2014b,miller2019}. A series of short-lived QPOs with center frequencies of $\sim$ 10-600\,Hz were found during the giant flares of SGR~1806$-$20 and SGR~1900+14 \citep{israel2005,strohmayer2005,strohmayer2006,huppenkothen2014b,miller2019}. Recently, \cite{2021Natur.600..621C} reported two broad high-frequency QPOs in the main peak of an extra-galactic giant flare lasting for \SI{3.5}{\milli\second}. During the short bursts from SGR~J1550$-$5418, which are in the same category with those from \sgrnos, a weak QPO signal at $\sim$ 260\,Hz was possibly found in a single burst \citep{2014ApJ...795..114H}. Using the same data but stacking some short bursts together, more significant QPOs centered at $\sim$93\,Hz and $\sim$ 127\,Hz, which are close to those found in giant flares, were reported \citep{2014ApJ...795..114H}. Also, a $\sim$57\,Hz QPO was found in stacked short bursts from SGR~1806$-$20 \citep{huppenkothen2014c}. All these QPOs can be explained by the oscillations in the crustal movement of magnetars \citep{huppenkothen2014b,miller2019}. These signals, though statistically not significant and also not common for most magnetar X-ray bursts, are among the few links between observations and physics of neutron stars.

Searching for QPOs in transient light curves requires special cares. Standard methods involving Fourier analysis are defined for stationary processes.
Magnetar bursts require special cares when performing Fourier analysis on their light curves because they have a beginning and end in duration and they are non-stationary processes. For stationary processes, the observed Leahy-normalized power should follow a $\chi_2^2$ distribution scaled by the mean power in each bin, but for the non-stationary signal the distribution at low frequency (<30 Hz) does not satisfy the $\chi_2^2$ distribution \citep{2013ApJ...768...87H,2014ApJ...795..114H}. In this paper, the periodogram is the squared modulus of the Fourier transform of the light curve. For transient events where the shape of the burst is known, Monte Carlo simulations of light curves can be obtained by sampling the fitted profile through a Poisson distribution \citep{2001MNRAS.321..776F, 2011MNRAS.415.3561G}. Each light curve can be used to generate a periodogram, thus one can get the distribution of the periodogram at different Fourier frequencies. The periodogram distribution at each frequency is a non-central $\chi_2^2(\lambda)$ distribution, whose non-central value $\lambda$ at each frequency is the power spectrum of the burst profile. This has disadvantages when the burst profiles of magnetars exhibit a variety of shapes, and thus the profiles could not be determined by the prior knowledge of the physical origin. This raises the possibility of false QPO detection or overestimating the significance of the potential QPOs.

There are other methods to search for QPOs in transient time series. A conservative approach is to model the periodograms by a Bayesian method \citep{2013ApJ...768...87H, 2010MNRAS.402..307V}.
We therefore search for the QPO signals using the Bayesian method during the short X-ray burst associated with FRB~200428, observed by the Hard X-ray Modulation Telescope (\insight{}) \citep{2021NatAs...5..378L} and the light curve of Konus-Wind \citep{2021NatAs...5..372R}. A significant QPO signal with center frequency at about 40\,Hz is detected with HXMT/ME using the Bayesian method and verified by the Konus-Wind data.

We structure the paper as follows. In section \ref{sec:instrument}, we present the details of the instrument and the data reduction process for this burst. The dead time correction of HXMT/ME is introduced as well. In section \ref{sec:method}, the result of the QPO search and its significance is presented using the data of HXMT/ME. The differences of the significance at different duration of the burst are also presented and discussed. The cross-correlation analysis between Konus-wind and HXMT/ME is also presented to verify the QPO. The physics interpretation of the QPO are discussed in section \ref{sec:physics}. We give a discussion and summary in section \ref{sec:summary}.

\section{The instrument and data reduction}
\label{sec:instrument}

\subsection{\insight{} /ME}
As the first X-ray astronomy satellite of China, \insight{} was launched on June 15, 2017 and carries three main instruments onboard \citep{2020SCPMAoverview}: the High Energy X-ray telescope (HE) \citep{2020SCPMAHE}, the Medium Energy X-ray telescope (ME) \citep{2020SCPMAME}, and the Low Energy X-ray telescope (LE) \citep{2020SCPMALE}.
The high flux of the X-ray burst associated with FRB~200428 caused the data saturation and loss in HE and LE detectors. The raw data in some time intervals were discarded on-board and their light curves have several gaps as described in \cite{2021NatAs...5..378L}. Therefore, only the data of the ME instrument is used to search for the QPO and ME also provides the high time resolution (6\,us) required for timing analysis \citep{2021arXiv210904709T}.

The data reduction of the X-ray burst associated with FRB~200428 is performed with the \insight{} Data Analysis Software package (\texttt{HXMTDAS}) version 2.04. The main steps of ME data reduction are: (1) Use the commands \texttt{mepical} in \texttt{HXMTDAS} to calibrate the photon events from the 1L data product according to the Calibration Database (CALDB) of \insight{}. (2) Select the good time intervals (GTIs) from $T_0$ to $T_0$+1\,s, where $T_0$ is 2020-04-28 14:34:24 UTC (Satellite time). (3) Identify the grade of each event according to their arrival time and calculate the dead time of each FPGA at the specified interval. (4) Extract the good events based on the GTIs using the commands \texttt{mescreen}. (5) Produce the light curve of ME with the dead time correction using the commands, \texttt{melcgen}. Although ME detectors do not have the problems of the data saturation like HE and LE, the dead time effect is still significant for bright X-ray burst detection.

We elaborate the dead time effect of ME instrument here. ME consists of three detector boxes. Each box has three independent Field Programmable Gate Array\,(FPGA) and each FPGA manages six Application Specified Integrated Circuit\,(ASIC). Each ASIC is responsible for the readout of 32 pixels \citep{2020JHEAp..27...64L}.
The dead time of ME can be calculated by each FPGA independently. When one pixel in the FPGA is triggered, a window with \SI{62.5}{\nano\second} is opened and the pixels triggered in other ASICs of the same FPGA will be recorded. If there are $N_{\rm{hit}}$ of ASICs in the same FPGA that are triggered within the time window, the electronics will take $163+93\cdot N_{\rm{hit}}$ microseconds to process these triggered pixels. The typical timescale of dead time is \SI{256}{\micro\second} because only one pixel is triggered at most situations \citep{2020SCPMAME}. The dead time process of ME is similar to that of non-paralyzed detectors, and the hardware algorithm is implemented in the \texttt{HXMTDAS} to calculate the dead time of ME. The dead time causes the loss of photons and the distribution of photon arrivals to deviate from Poisson distribution, and alters the distribution of power in the periodogram as well. Note that the dead time strongly depends on the count rate. As a consequence, dead time corrections are especially important for bright bursts.

\begin{figure}
\begin{center}
\includegraphics[scale=0.5]{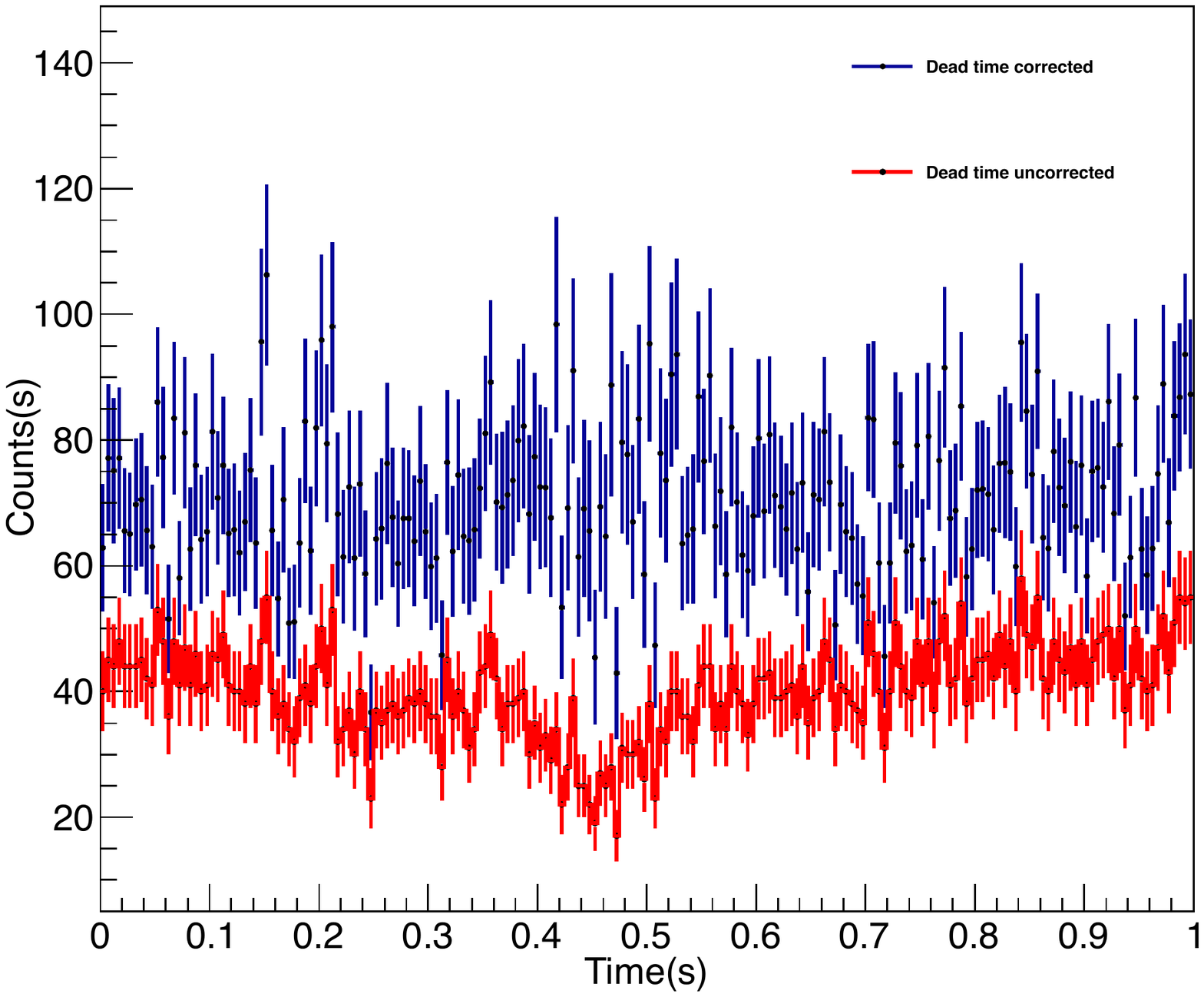}
\caption{
Light curve of the special events obtained by \insight{}/ME near the FRB~200428 burst, where time zero is 2020-04-28 14:34:24 UTC (Satellite time). The blue line is the light curve after dead time correction and the red line is the light curve without dead time correction. After dead time correction, the counts of these special events keep stable.
\label{fig_MElcdeadtime}}
\end{center}
\end{figure}

A special type of events in ME data could be an indicator to determine whether the dead time correction is performed well. There are two kinds of thresholds for ME. One is the triggered threshold, the other is the readout threshold. The special events are those the pulse height is higher than the triggered threshold and lower than the readout threshold.
These events are mainly the noise events which are independent of the source brightness, so the light curve of these events is expected to be constant after dead time correction as shown in Figure \ref{fig_MElcdeadtime}.

\subsection{Konus-Wind}
Konus-Wind (KW) consists of two identical NaI(Tl) scintillation detectors, each with $2\pi$ sr field of view. One detector (S1) points towards the south ecliptic pole, thereby observing the south ecliptic hemisphere, while the other (S2) observes the north ecliptic hemisphere. Here the light curves from KW are obtained from \cite{2021NatAs...5..372R}.

\subsection{Methods for the QPO analysis}

\subsubsection{Monte Carlo simulations of pulse profile}
\label{subMC}
Monte Carlo simulation of light curves is a standard tool in timing analysis.
One can fit an empirical function to the burst profile and then
generate a large number of realizations of that burst profile by adding the Poisson photon counting noise. The periodograms computed from these simulated light curves form a distribution against which to compare the periodogram of the real data.
The X-ray burst profile of FRB~200428 are fitted with three Gaussian functions, which can be found in \cite{2021NatAs...5..378L},
\begin{equation}
R_{3}=\sum_{i=1}^{3}{N_{{\rm p},i}G(t,t_{{\rm p},i},\sigma_{{\rm p},i})}+l,
\label{eq0}
\end{equation}
where $G(t,t_{{\rm p},i},\sigma_{{\rm p},i})=\frac{1}{\sqrt{2\pi}\sigma_{{\rm p},i}}\exp(-\frac{(t-t_{{\rm p},i})^2}{2\sigma_{{\rm p},i}^{2}})$, $N_{{\rm p},i}$, $t_{{\rm p},i}$ and $\sigma_{{\rm p},i}$ are the normalisation, arrival time and Gaussian width of the $i$th peak. $l$ is the background level of the light curve.

Actually, the true pulse profile of the X-ray burst associated with FRB~200428 can not be known precisely due to the degeneracy between the overall burst profile and the red noise component. If the pulse profile can be presented by eq. (\ref{eq0}), Monte Carlo simulations of pulse profile caused by the Poisson noise of photon counting can be used to estimate the significance of the QPO candidate.

\subsubsection{Bayesian method}
\label{subsectionBaymethod}
In the following, we briefly introduce the Fourier-based periodograms analysed by Bayesian method commonly used to derive the significance of QPOs \citep{2010MNRAS.402..307V, 2013ApJ...768...87H}.
A periodogram is a sample of the power spectral density (PSD) of the signal based on a given time series $x_k(t)\, (k=0,1,...N-1), N=2^m$ (m is an integer). We calculate the periodogram as the absolute square of the discrete Fourier transform of the signal, and it follows $\chi_2^2$ distribution around the PSD $S_j = S(f_j), j=0,1, N/2$ at Fourier frequency $f_j$ \citep{van1989fourier,1995A&A...300..707T,2010MNRAS.402..307V}. Therefore, the periodogram $I_j$ follows an exponential distribution about $S_j$ and can be described below,
\begin{equation}
    \label{eq:periodogram_exp}
    p(I_j \mid S_j) = \frac{1}{S_j}e^{-I_j/S_j}.
\end{equation}

Many astrophysical transients show excess power at low frequencies, and it is often assumed that this can be modeled as a red noise process. One basic noise model $S_p(f)$ that is commonly used is a combination of a red noise power law and a white noise amplitude,
\begin{equation}
    \label{eq:noisemodel1}
    S_p(f) = Af^{-\alpha} + C,
\end{equation}
where $A$ is the amplitude and $\alpha$ is the index. $C$ is the constant representing the white noise component. Another noise model $S_b(f)$ is composed of broken power law and a constant white noise,


\begin{equation}
 \label{eq:noisemodel2}
S_b(f) = Bf^{-\alpha_1} +C,  (f \leq f_{break}) 
\end{equation}  

\begin{equation}
S_b(f) = Bf_{\rm{break}}^{\alpha_2 - \alpha_1}(f)^{-\alpha_2} +C, (f>f_{break})
\end{equation}

where $\alpha_1$ and $\alpha_2$ are indices of the two power law components, and $f_{\rm{break}}$ is the break point for the frequency, and $B$ is the normalization value. When $\alpha_1$ equals $\alpha_2$, the model of $S_b(f)$ will be the same as $S_p(f)$ and they satisfy the requirement of nested models.
Following Bayes' rule, the posterior probability of a set of model parameters $\theta$,  like $A$, $\alpha$ and $C$ in noise model $S_p(f)$, can be defined given the uniform distribution of the prior.

We use the Bayesian method to maximum likelihood estimation (MLE) to the observed Leahy-normalized periodogram and obtain the maximum a posterior (MAP) estimates of the model parameters.
As introduced in \cite{2010MNRAS.402..307V}, maximizing the joint likelihood function $p(\bm{I} | \bm{\theta}, H) = \prod\limits_{j}^{N/2}p(I_j | S_j)$  is equivalent to minimizing the deviance function:
\begin{equation}
    \label{eq:deviance}
    D(\bm{I}, \bm{\theta}, H) = -2 \log p(\bm{I} \mid \bm{\theta}, H) = 2 \sum\limits_{j=1}^{N/2}(\frac{I_j}{S_j} + \log S_j).
\end{equation}

In other words, finding the maximum likelihood estimates (MLEs) of the parameters of a model $S(\bm{\theta})$ can be expressed as
\begin{equation}
    \label{eq:MLE}
    \bm{\hat{\theta}_{\rm{MLE}}} =  \arg \min_{\theta} D(\bm{I^{\rm{obs}}}, \bm{\theta}, H)
    = \arg \max_{\theta} p(\bm{I} = \bm{I^{\rm{obs}}} \mid \bm{\theta}, H),
\end{equation}
where $\bm{I^{\rm{obs}}}$ is the observed periodogram, $\arg \min$ is to find the set of parameter $\bm{\theta}$ to obtain the minimum value of function $D$, and $\arg \max$ of distribution $p$ is to find the maximum likelihood.

To verify the possible QPO in the periodogram, a model selection task is taken to assess whether the periodogram can be represented by a simple power law or it requires a more complex model like broken power law.
The likelihood ratio test (LRT) is employed to distinguish two nested model, \textit{null hypothesis} $H_0$ and the \textit{alternative hypothesis} $H_1$. $H_0$ is the power law and $H_1$ is the broken power law. The LRT statistic is twice the minimum log likelihood of the two models:
\begin{equation}
    \label{eq:lrt}
    T_{\rm{LRT}} =  -2 \log \frac{p \left( \bm{I} \mid \bm{\hat{\theta}^{0}_{\rm{MLE}}}, H_0 \right)}{p \left( \bm{I} \mid \bm{\hat{\theta}^{1}_{\rm{MLE}}}, H_1 \right)}
    =  D_{\rm{min}}(H_0) - D_{\rm{min}}(H_1).
\end{equation}

Once the noise model is selected using the LRT statistic, the QPO search can also be considered as a model selection problem. The new $H_1$ model is the sum of the selected noise model and a Lorentz model. When the amplitude of the Lorentz is zero, the $H_1$ model becomes the noise model.
The asymptotic theory suggests that the LRT statistic should be distributed as a chi-square variable under certain regularity conditions, $T_{\rm{LRT}} \sim \chi_\nu^{2}$, where the number of degrees of freedom $\nu$ is the difference between the number of free parameters in $H_1$ and $H_0$. When the regularity conditions are not satisfied, one can not use the distribution to be that of the asymptotic theory \citep{2002ApJ...571..545P}. Nevertheless, LRT is still a powerful statistic for comparing models and can be calibrated by posterior predictive simulations, as shown by \cite{2002ApJ...571..545P}.

In order to investigate narrow features, another statistic $T_{\rm{R}}$ to use is the maximum ratio of observed to model power \citep{2010MNRAS.402..307V, 2013ApJ...768...87H},
\begin{equation}
    \label{eq:TR}
    T_{\rm{R}} = {\rm{max}} (R_j),
\end{equation}
where
\begin{equation}
    \label{eq:Rj}
    R_j = 2I_j/R_j,
\end{equation}
$I_j$ and $S_j$ are the observed and model powers. This is the candidate for single bin periodicity when searching the periodogram for the highest data/model outlier. After simulating a large number of periodograms from the posterior parameter sets and fitting each simulated periodogram with the preferred noise model, we can also compute the data/model and find the maximum data/model outlier. Then the distributions of the highest outlier can be calibrated under the selected noise model.
The posterior p-value for the observed outlier can be obtained when calculating the tail areas from the simulated outlier distributions.

\section{Results}
\label{sec:method}
After we have obtained the dead time corrected light curve of ME in 18--50\,keV as displayed in Figure \ref{fig_MElconly}, the periodograms of this burst is calculated using the Fast Fourier Transform (FFT) code implemented in open source Python software \texttt{stingray} \citep{2019ApJ...881...39H}. We find a potential QPO candidate with moderate significance at frequency about 40\,Hz. We use two different strategies to evaluate the significance of the potential QPO signal: (1) Monte Carlo simulation; (2) Bayesian approach. Then the significance at different duration is also investigated. The cross-correlation between ME and KW light curves is performed, which also verifies the QPO signal at about 40\,Hz.

\subsection{Monte Carlo simulations of pulse profile}
\label{sec:monte carlo}
Although Monte Carlo simulation of the pulse profile is an important tool in timing analysis, it must know the physically motivated burst profile function in advance. We assume the burst profile of the X-ray burst associated with FRB~200428 can be fitted with three Gaussian functions and the fit parameters can be found in Table \ref{tab:spec_parameters}.
It should be noted that we just take the burst as an example to show an assumption for the burst profile without taking the red noise into account will produce an uncorrected significance estimation of the QPO candidate.

\begin{figure}
\begin{center}
\includegraphics[scale=0.75]{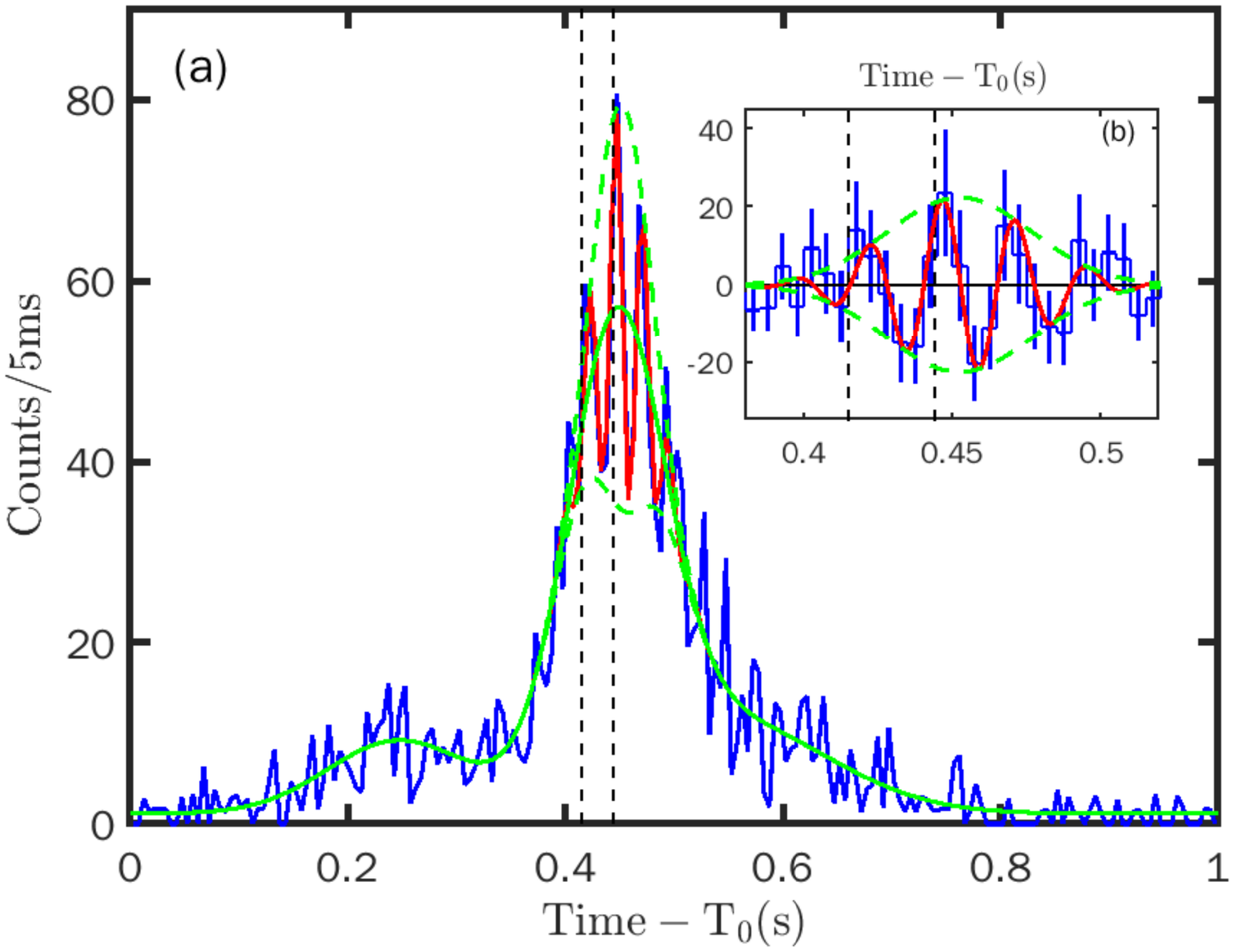}
\caption{
Light curve of the X-ray burst in 18--50\,keV obtained by \insight{} /ME with time resolution of 5\,ms. The blue line is the light curve after dead time correction. The green line represents the fit result of the burst profile with three Gaussian functions as given in eq. (\ref{eq0}). The red line is the fitted oscillation curve with eq. (\ref{eq2}). The two dashed green lines show the lower and upper envelopes of the amplitude variation of the QPOs described by eq. (\ref{eq_en}). The two vertical dashed lines denote the times of the two radio pulses of FRB~200428. The inset is the light curve around the burst peak with the solid green line subtracted.
\label{fig_MElconly}}
\end{center}
\end{figure}

\begin{figure}
\begin{center}
\includegraphics[width=0.6\textwidth, angle=90]{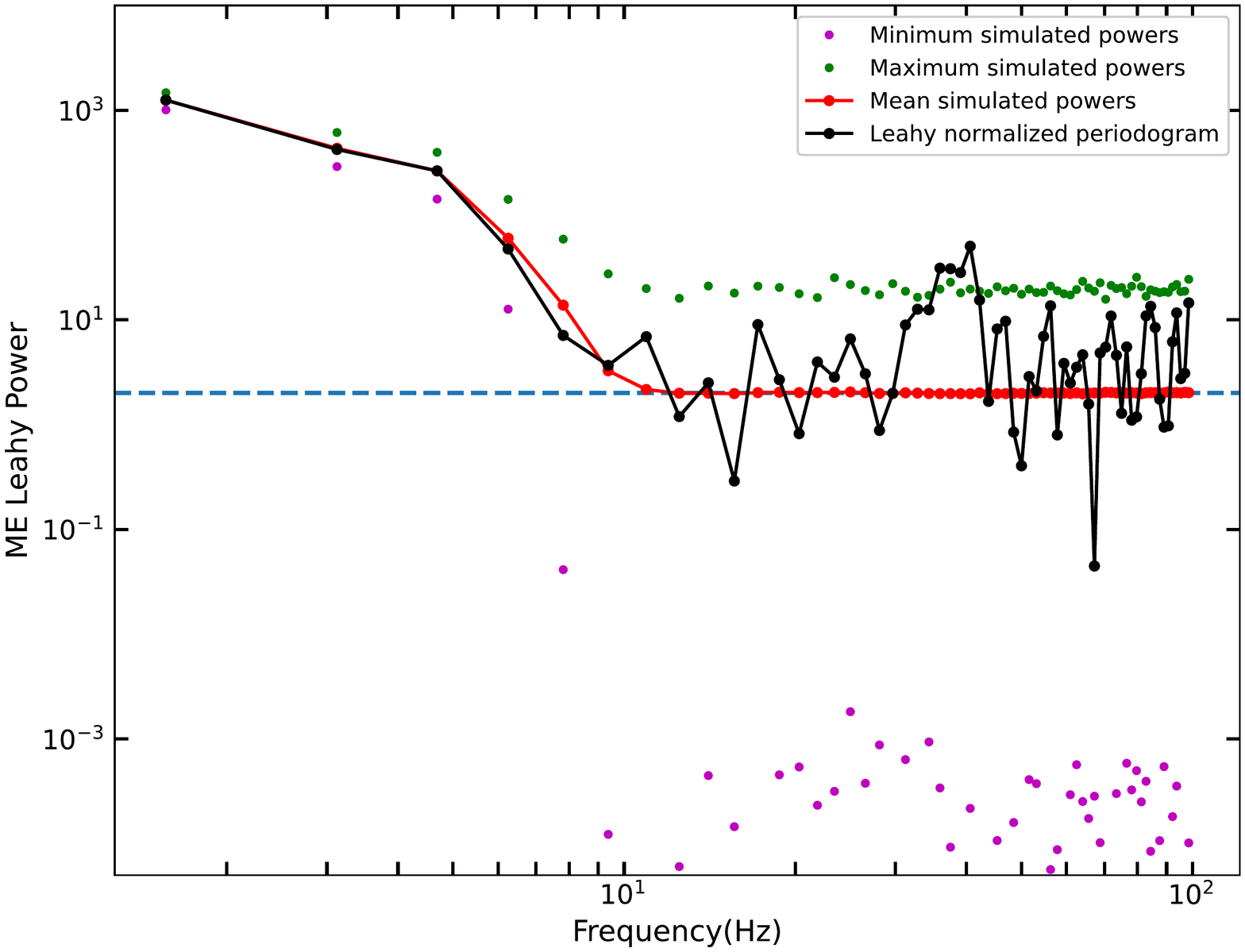}\caption{
The Leahy-normalized periodograms of the light curves of \insight{}/ME with time resolution of 5\,ms. The solid black line represents the Leahy-normalized periodograms.
10000 simulated light curves sampled from the fitted three-Gaussian burst profile by adding the Poisson noise is used to generate the Leahy-normalized periodograms in the same way as we did for the data.
Thus the distribution of Leahy power in each frequency bin can be derived. The mean simulated power in each frequency bin is shown in red solid line with the maximum (green dots) and minimum (magenta dots). The blue dotted line means the Leahy power with 2.
\label{fig_mepds}}
\end{center}
\end{figure}

The procedures of the Monte Carlo simulations of the burst profile is described below.
As introduced in section \ref{subMC}, the burst profile can be fitted with eq. (\ref{eq0}) and shown in Figure \ref{fig_MElconly} with green solid line. The Leahy-normalized periodograms of light curve can be derived using the Fast Fourier Transform (FFT) code in Python software \texttt{stingray} as shown in Figure \ref{fig_mepds}. A potential QPO candidate with moderate significance at frequency about 40\,Hz can be found. In order to establish the p-value of the QPO, Monte-Carlo simulations of the burst profile are used to validate whether the QPO signal is generated from the fluctuations of the light curve or not.
We then generate 10000 simulated light curves sampled from the fitted three Gaussian burst profile by adding the Poisson photon counting noise. The periodogram is computed for each simulated light curve to form a distribution at each frequency bin. The maximum, mean and minimum values of power distribution at each frequency bin are calculated and plotted in Figure \ref{fig_mepds}.
Comparing the observed power in each frequency bin with the distribution of simulated powers allows us to make a statement that the observed power at 40\,Hz is difficult to be produced by the Poisson noise process alone according to the distribution of simulated powers.
From the simulated distribution at each frequency, the distribution of Leahy power above 10\,Hz satisfies a $\chi^2_2$ distribution. Consequently, the p-value for the QPO at 40\,Hz under the assumption of the three-Gaussian burst profile is 1.7e-7, indicating that the Poisson noise of the profile with only three broad Gaussians cannot explain the observed excess in power at 40\,Hz.

The QPO candidate in the light curve as shown in Figure \ref{fig_MElconly} can also be fitted with,
\begin{equation}
Q=N_{\rm g}G(t,t_{\rm g},\sigma_{\rm g})\sin(2\pi{f_{\rm q}(t-t_{\rm q})})+R_{\rm 3},
\label{eq2}
\end{equation}
where $N_{\rm g}$, $t_{\rm g}$ and $\sigma_{\rm g}$ are the normalization, arrival time and width of the broad Gaussian profile of the X-ray burst. $f_{\rm q}$ is QPO frequency and $t_{\rm q}$ is the time offset of sinusoidal function. The fitting results are $N_{\rm g}=1.4\pm0.2$, $t_{\rm g}=0.452\pm0.004$, $\sigma_{\rm g}=0.024\pm0.004$, $f_{\rm q}=40.4\pm1.2$, $t_{\rm q}=0.09\pm0.01$. The frequency $f_{\rm q}$ is consistent with the result from the periodogram analysis presented below. The lower and upper envelopes of the amplitude modulation of the QPO candidate can thus be expressed by
\begin{equation}
Q_{\rm en}=R_{\rm 3}\pm{N_{\rm g}G(t,t_{\rm g},\sigma_{\rm g})}.
\label{eq_en}
\end{equation}
The fitted ME light curve with eq. (\ref{eq2}) is shown in Figure \ref{fig_MElconly} (b), where the lower and upper envelopes of the amplitude modulation of the QPO described by eq. (\ref{eq_en}) are also plotted.

\begin{table}
\footnotesize
\caption{Fitting parameters of the light curve of \insight{}/ME\,(18-50\,keV) with eq.(1).}
\scriptsize{}
\label{tab:spec_parameters}
\medskip
\begin{center}
\begin{tabular}{c c c c cc c c c c c c}
\hline \hline
Inst.& $N_{\rm p1}$ & $t_{\rm p1}$ & $\sigma_{\rm p1}$ & $N_{\rm p2}$& $t_{\rm p2}$ & $\sigma_{\rm p2}$& $N_{\rm p3}$& $t_{\rm p3}$ & $\sigma_{\rm p3}$& $l$ & \\
\hline
ME & 49.7& 0.446 & 0.044 & 10.4 &0.545 & 0.095 &8.02 & 0.246 & 0.065 & 1.129 \\
\hline \hline
\end{tabular}
\end{center}
\end{table}

\subsection{Bayesian method}
The Monte Carlo method outlined above is versatile and powerful, but it has limitations.
In this section, we will explore the Bayesian method which is a conservative method based on the assumption that red noise component dominates the periodogram. This approach is to model the observed periodogram rather than the light curve.
The data between 0.1\,s to 0.74\,s is used to generate the dead time corrected light curve with time resolution of 2.5\,ms and the corresponding periodogram. The 0.64\,s duration basically covers the entire X-ray burst of FRB~200428 as shown in Figure \ref{fig_MElconly}.
In order to study the non-stationary effect on this burst, we also compare the results of different duration time such as the main peak of the burst (0.32\,s) and more background data besides the burst such as 1.28\,s.

\subsubsection{Selection of a noise model}
\begin{figure}
\begin{center}
\includegraphics[scale=0.6]{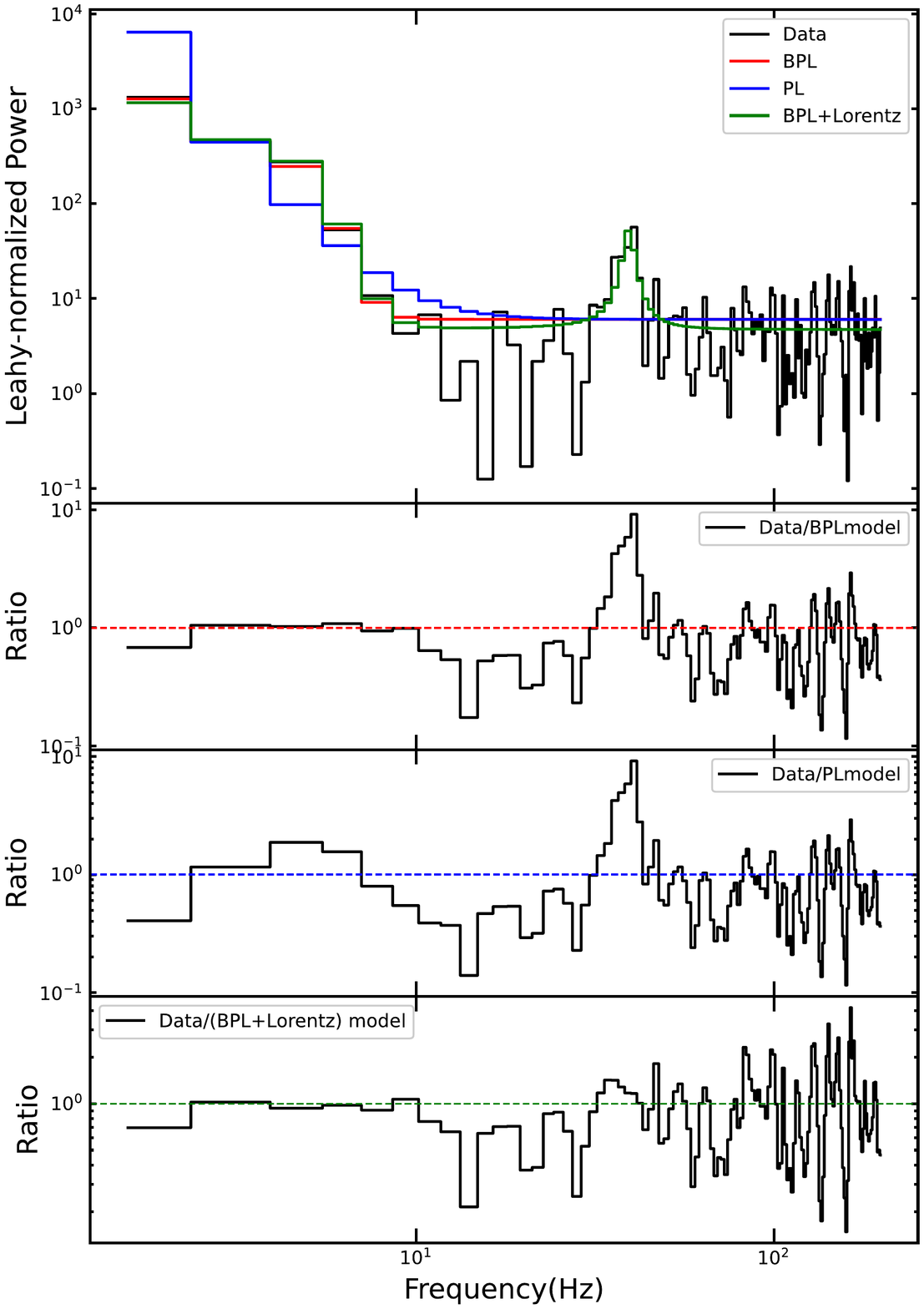}
\caption{
The Leahy-normalized periodogram and residuals of the different models. Upper panel: The Leahy-normalized periodogram (black), power-law fit (blue), broken power law fit (red) and broken power law plus the Lorentz fit (green) of the periodogram. The three lower panels show the smoothed ratio of the observed periodogram and power law fit, broken power law fit, and broken power law plus Lorentz fit with MAP estimates, respectively. The smooth is derived by a Wiener filter with a width of three frequency bins in order to reduce the probability of the minimization algorithm terminating in local minima due to sharp noise features. The QPO candidate with center frequency of 40 Hz is clearly shown in the ratio plot.
\label{fig_BPLandPL}}
\end{center}
\end{figure}

In the following, we briefly illustrate the analysis procedure with the X-ray burst associated with FRB~200428. This burst has a duration of $T_{90}$ with 0.65\,s and the periodogram of 0.64\,s (from 0.1 to 0.74\,s) length with time resolution of 2.5\,ms for this burst is presented in Figure \ref{fig_BPLandPL}.
After fitting both a simple power law\,($H_0$ model) and the broken power law\,($H_1$ model), the likelihood ratio between the two models using eq.(\ref{eq:lrt}) is calculated as LRT = 5.36. The fits to the periodogram and the ratios (data/model) are presented in Figure \ref{fig_BPLandPL} using the MAP estimate of the parameters.

\begin{figure}
\begin{center}
\includegraphics[scale=0.5]{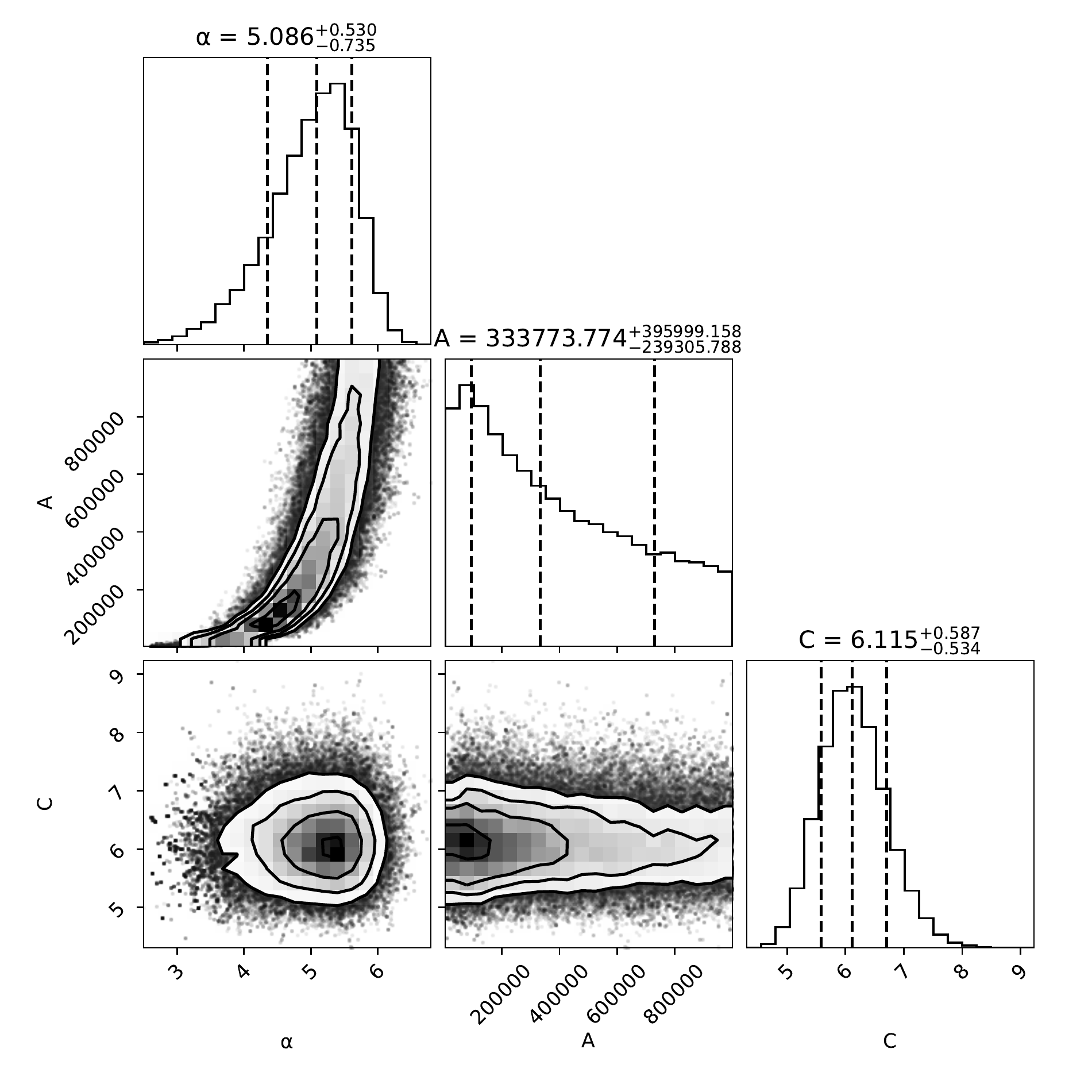}
\caption{
The posterior distributions of parameters in the simple power law are presented on the diagonal. The parameter $A$ is the amplitude and $\alpha$ is the index of the power law as defined in eq.(\ref{eq:noisemodel1}). $C$ is the constant representing the white noise component and also defined in eq.(\ref{eq:noisemodel1}). The off-diagonal panels show correlations between parameters. The scatter plots are selected from the last 10000 picked parameter pairs from the entire sample of 100000 parameter sets. We can observe that a very tight correlation between power law index and normalization, and very little correlation between the normalization and the white noise.
\label{fig_PLposD}}
\end{center}
\end{figure}

\begin{figure}
\begin{center}
\includegraphics[scale=0.6]{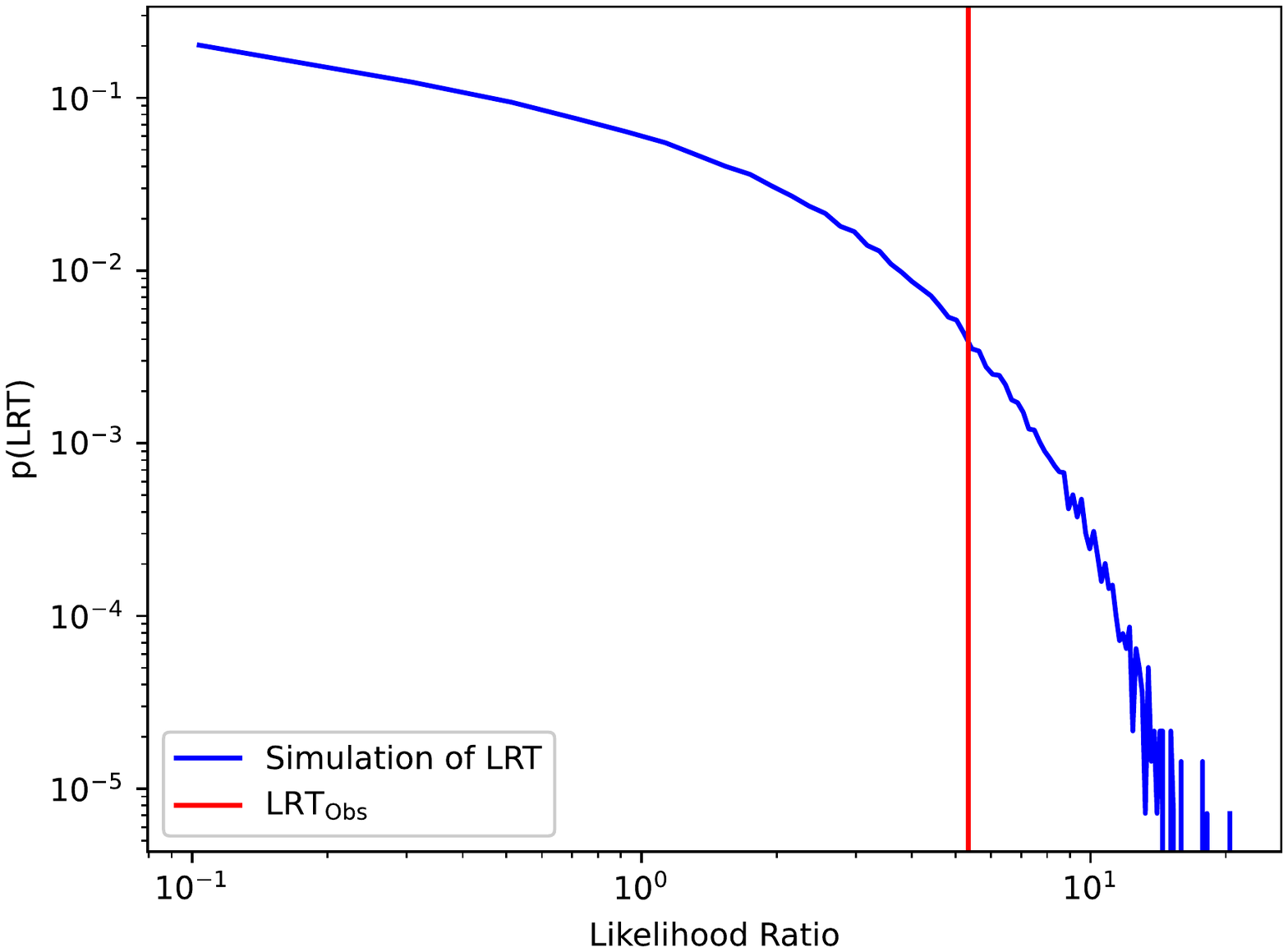}
\caption{
Distribution of likelihood ratios for 150000 simulations of the null hypothesis (power-law model). The observed value of LRT is indicated as the red vertical line. The ratio of the tail area is about 0.03, indicating that a more complex model (in this case the broken power law) may be more appropriate in modeling the broadband noise.
\label{fig_PLRT}}
\end{center}
\end{figure}

We utilize the Markov Chain Monte Carlo simulations (MCMCs) implemented in Python package \texttt{emcee} \citep{foreman2013emcee} to generate sets of parameters of posterior probabilities. We then sample from the posterior distribution of the power law model via MCMC and simulate the periodograms according to eq. (\ref{eq:periodogram_exp}) from chains of posterior distribution as shown in Figure \ref{fig_PLposD}.
These simulated periodograms are again fitted with models $H_0$ and $H_1$, respectively, so that we can build the likelihood ratio distribution of the power law model. Comparing the distribution of simulated likelihood ratios with that of the observed LRT, we can compute the tail area probability or the posterior p-value as 0.03 which is shown in Figure \ref{fig_PLRT}. Thus we can reject the power law model and accept the broken power law model for the noise model.

\subsubsection{Searching for QPO}
The noise model has been selected above as broken power law, then we can search for the QPO as a model selection problem. We compare the selected noise model to a more complex model combining both the noise model and a Lorentz function to account for the QPO.
The burst periodogram are fitted by \texttt{bknpow+constant} as a simple model ($H_0$), and can be also fitted by \texttt{bknpow+Lorentz+constant} as a nested model ($H_1$). \texttt{bknpow} and \texttt{Lorentz} are the power law function with a break frequency and the Lorentz function in \texttt{Xspec}. After fitting the periodogram with the two models above, the likelihood ratio between the two model can be obtained, denoted as $T_{\rm{LRT}}^{\rm{obs}}$.
The $D_{\rm{min}}(H_0)$ and $D_{\rm{min}}(H_1)$ of the observed burst periodogram are 743.75 and 712.76, respectively. The ratio of the periodogram and $H_1$ model is also displayed in the bottom panel of Figure \ref{fig_BPLandPL}. Thus the observed LRT of the burst is $T_{\rm{LRT}}^{\rm{obs}} = 30.99$.

Next, we will clarify how we calibrate the posterior distribution of LRT and calculate the p-value. Using the $H_0$ model, we pick $n$ sets of parameters for the $H_0$ model from the posterior MCMC chains. Each set of parameters is simulated to produce a periodogram, for which an LRT value for $H_0$ and $H_1$ is computed. We then have $n$ LRT values for $n$ periodogram. The distribution of the $n$ LRT values is the posterior distribution of LRT which is shown in Figure \ref{fig_allLRT} with red solid line.
The observed LRT value with red vertical line is also shown in Figure \ref{fig_allLRT}. As a result, the p-value or the ratio of the tail areas of $T_{\rm{LRT}}^{\rm{obs}}$ is 1e-6, which corresponds to 4.75\,$\sigma$.
We find a posterior centroid frequency for the QPO of $39.3_{-1.20}^{+1.22}$ Hz, with a width of $3.7_{-2.01}^{+3.74}$ Hz. We use the posterior distribution of QPO parameters to derive the fractional root-mean-squared (RMS) amplitude $A_{frac}$, and find $A_{frac}= 0.36 \pm{0.096}$. The results of 0.64\,s duration is also displayed in Table \ref{tab:LRT}.

\subsection{Significance at different duration}
The analysis presented above makes a strong assumption about the data: our choice of a $\chi^2_2$ distributed likelihood around the model power spectrum implies that the periodogram is the result of a stationary noise process.
This assumption is not strictly the case for burst signals, as shown in \cite{2013ApJ...768...87H}, it is a conservative assumption that holds for all but the lowest frequencies in the periodogram.
In this paper, we also explore the effects of the non-stationary nature of a transient light curve, and a QPO within it.
We show that selecting more background signals before or after the burst signal leads to greatly overestimating the significance of potential QPOs. We also show that the significance will be overestimated if the whole burst signal is selected for analysis but the QPOs occur in only part of the burst.

\begin{figure}
\begin{center}
\includegraphics[scale=0.45, angle=90]{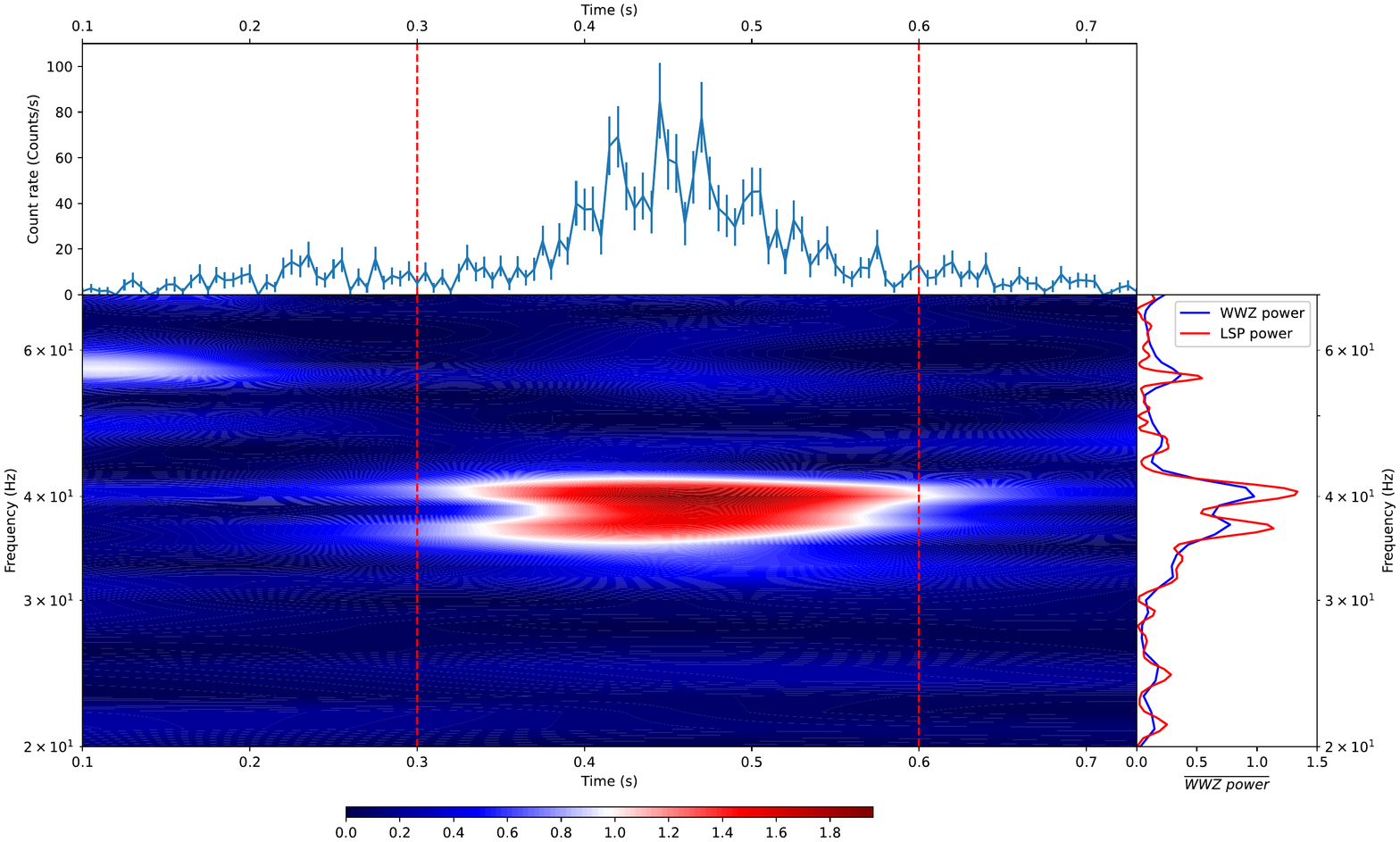}
\caption{
Wavelet analysis of the ME light curve from 0.1\,s to 0.74\,s with time resolution of 2.5\,ms. The top panel is the light curve and the left bottom panel is the result of the wavelet analysis, the 40\,Hz QPO occurs in the time range of 0.3\,s to 0.6\,s, which is indicated with red dotted lines. The right bottom panel represents the WWZ power and LSP power.
\label{fig_wavelet}}
\end{center}
\end{figure}

To search for the duration of the 40\,Hz QPO, we utilize wavelet analysis and two most widely used methods: the generalized Lomb-Scargle Periodogram (LSP; \cite{1976Ap&SS..39..447L, 1982ApJ...263..835S}) and Weighted Wavelet Z-transform (WWZ; \cite{1996AJ....111..541F}) are used to obtain the power spectra of the 0.64\,s light curve with time resolution of 2.5\,ms. In this work the power spectra of LSP method is checked with the independent results of WWZ approach. As shown in Figure \ref{fig_wavelet}, the QPO signal appears in the main peak of time range from 0.3\,s to 0.6\,s. Then we choose the duration of the main peak about 0.32\,s and more background before and after the main peak, such as 1.28\,s to compare their differences.

\begin{figure}
\begin{center}
\includegraphics[scale=0.5]{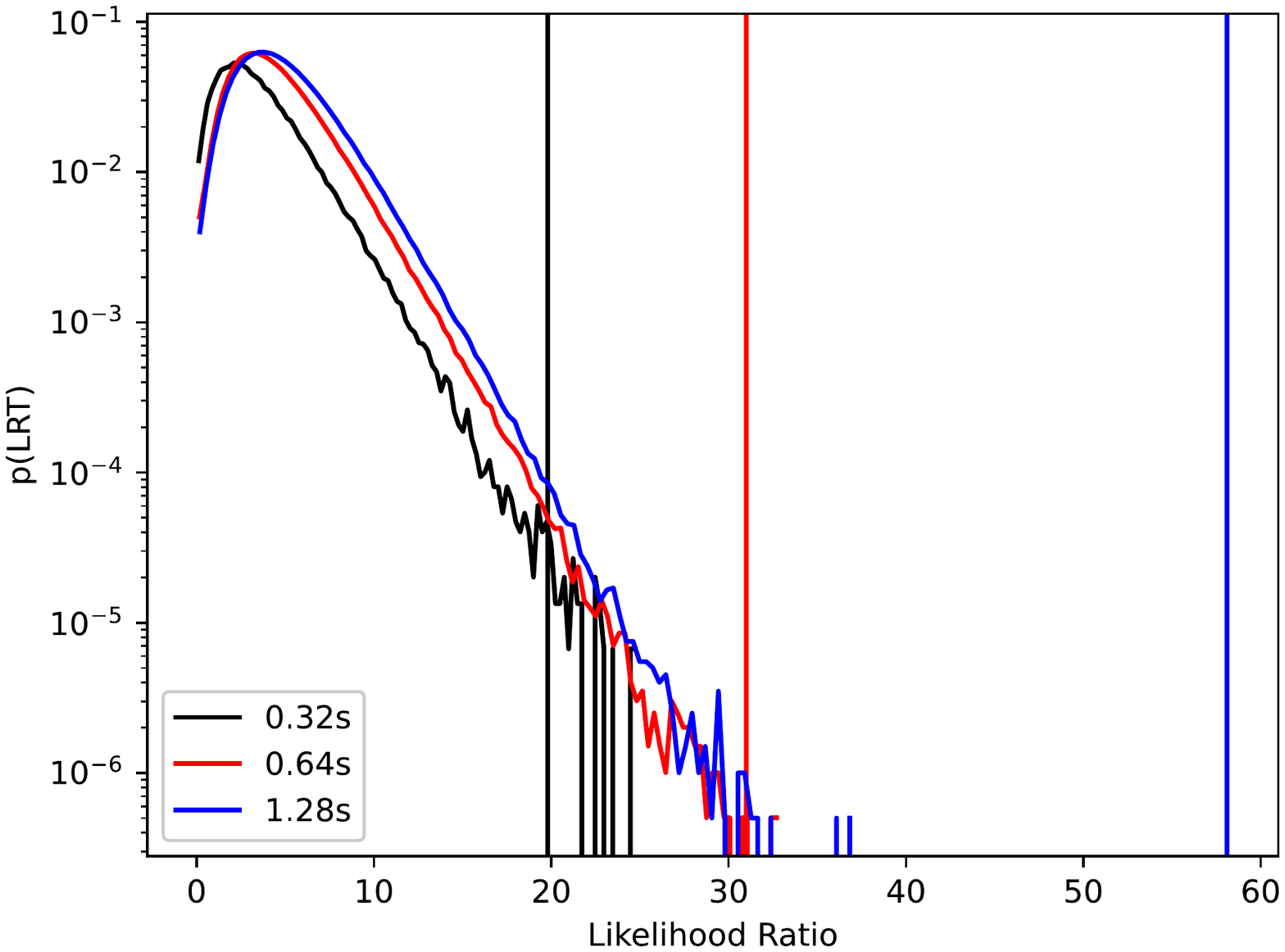}
\caption{
Distributions of likelihood ratio test for 150000, 2000000 and 2000000 simulations of the null hypothesis for 0.32\,s (black), 0.64\,s (red) and 1.28\,s (blue), respectively. The corresponding $H_0$ model is power law, broken power law and broken power law, respectively. The observed values of LRT are indicated as vertical lines. The ratios of the tail area for 0.32\,s and 0.64\,s are 2.9e-4 and 1e-6, both are indicating that a more complex model including additional Lorentz model may be more appropriate in modeling the periodogram. As for the 1.28\,s duration, the tail area is zero and the p-value is an upper limit.
\label{fig_allLRT}}
\end{center}
\end{figure}

Using the same method for the 0.64\,s, we select the noise model for 0.32\,s and 1.28\,s.
$H_0$ model is power law for 0.32\,s and broken power law for 1.28\,s.
As shown in Table \ref{tab:LRT}, for each duration time, after selecting the noise model $H_0$, we use $H_0$ model to generate a large sets of parameters (for power law, three parameters as shown in eq.(\ref{eq:noisemodel1}) and for broken power law, five parameters as shown in eq.(\ref{eq:noisemodel2})) from their posterior distributions. Each parameter set is used to simulate a periodogram using the exponential distribution in eq.(\ref{eq:periodogram_exp}). For the simulated periodogram we also use $H_0$ model and $H_1$ model to get the $D_{\rm{min}}$ values as defined in eq.(\ref{eq:deviance}) using MCMC. The $T_{\rm{LRT}}$ can be derived from eq.(\ref{eq:lrt}). Then the distribution of the simulated likelihood ratios can be derived from multiple sets of parameters of $H_0$ model. Comparing the distribution of simulated likelihood ratios with that of the observed $T_{\rm{LRT}}$ as shown in Figure \ref{fig_allLRT}, we can compute the tail area probability and this is also the normalized posterior p-value.
150000 and 2000000 sets of parameters for the $H_0$ model from the posterior Markov chain are used to simulate the periodogram for 0.32\,s and 1.28\,s, respectively.
The likelihood ratio distributions and the observed LRT values are also displayed in Figure \ref{fig_allLRT}.
The posterior parameters of the QPO, such as the centroid frequency, width and normalization are also shown in Table \ref{tab:LRT}.

The smaller p-value shown in Table \ref{tab:LRT} indicates a higher significance of the QPO. Although the parameters of the QPO are consistent within the uncertainties in the three durations, however, the uncertainties of the QPO parameters are smaller for the longer duration.
Our explanation about why the significance is higher for a long duration used is as follows. We find that the QPO is only present in the main peak of the X-ray burst. However, the noise level varies through the time series, i.e., the noise level is higher during the main peak where the count rate is higher (and also varies) than that outside the main peak. This is a typical non-stationary process. Extending the selected duration beyond the main peak is equivalent to using the noise level beyond the main peak to estimate the noise level during the main peak, which does not change the strength of the QPO but decreases the estimated noise. Consequently, the signal to noise ratio and detection significance the QPO are artificially increased.

To test this explanation, we have also done simulations for 0.64\,s and 1.28\,s and assume the light curve before and after the main peak (0.32\,s) are only Poisson processes, i.e., by replacing the observed light curve with the Poisson sampled light curve of the same count rate. We have obtained the same results: a longer duration outside the main peak leads to a higher significance of the QPO. Therefore, the reliable way to estimate the significance of the QPO is to use the segment of data in which the QPO is present.

As a consequence, we report the final results of QPO from 0.32\,s which are shown in Table \ref{tab:LRT}.

\begin{table}
\footnotesize
\caption{Results for the significance of QPO  and QPO parameters at different duration time using the LRT statistic.}
\scriptsize{}
\label{tab:LRT}
\medskip
\begin{center}
\begin{tabular}{c c c c c c c c c c}
\hline \hline
Duration time & $H_0$ model & $H_1$ model & $D_{min}(H_0)$ & $D_{min}(H_1)$ & $T_{\rm{LRT}}^{\rm{obs}}$ & p-value & Frequency & Width  & Norm \\
(s)  &  &  &  &  &  &    &  (Hz) & (Hz) & \\
\hline
 0.32 & PL  &  PL+Lorentz   & 378.91   &  359.07  & 19.84    &  2.9e-4 & $39.14_{-1.94}^{+1.88}$ & $4.88_{-2.73}^{+5.67}$ & $389.15_{-167.17}^{+260.55}$\\
\hline
 0.64 & BPL  &  BPL+Lorentz  & 743.75  & 712.76  & 30.99  & 1e-6 & $39.28_{-1.20}^{+1.22}$ &  $3.67_{-2.01}^{+3.74}$ &  $315.27_{-116.74}^{+218.63}$\\
\hline
 1.28 & BPL  &  BPL+Lorentz  & 1479.71 & 1421.63  & 58.08    & <5e-7 & $39.26_{-0.8}^{+0.78}$ & $3.24_{-1.41}^{+2.02}$  & $267.95_{-76.85}^{+135.87}$\\
\hline \hline
\end{tabular}
\end{center}
\end{table}

As described in the section \ref{subsectionBaymethod},
another statistics for investigating narrow QPOs is $T_{\rm{R}}$, which is defined in eq.(\ref{eq:TR}). We also search the periodogram for the highest data/model outlier and compare this outlier to those distributed by pure noise to find narrow features that may be candidates for a possible QPO.
One shortcoming of the $T_{\rm{R}}$ statistic is that it optimally detects periodic signals confined to one frequency bin. Since the QPO power spreads over several bins, this is not an optimal way of detecting broad signals. There are several ways to reduce this restriction. One is to bin (or smooth) the data in some way, and compute $T_{\rm{R}}$ for the binned data. If we bin the simulated periodograms in the same way, then the test statistic $T_{\rm{R}}$ for the binned data is comparable to the distribution approximated by our simulations, and the latter can be used to derive posterior predictive p-values.

\begin{figure}
\begin{center}
\includegraphics[scale=0.5, angle=90]{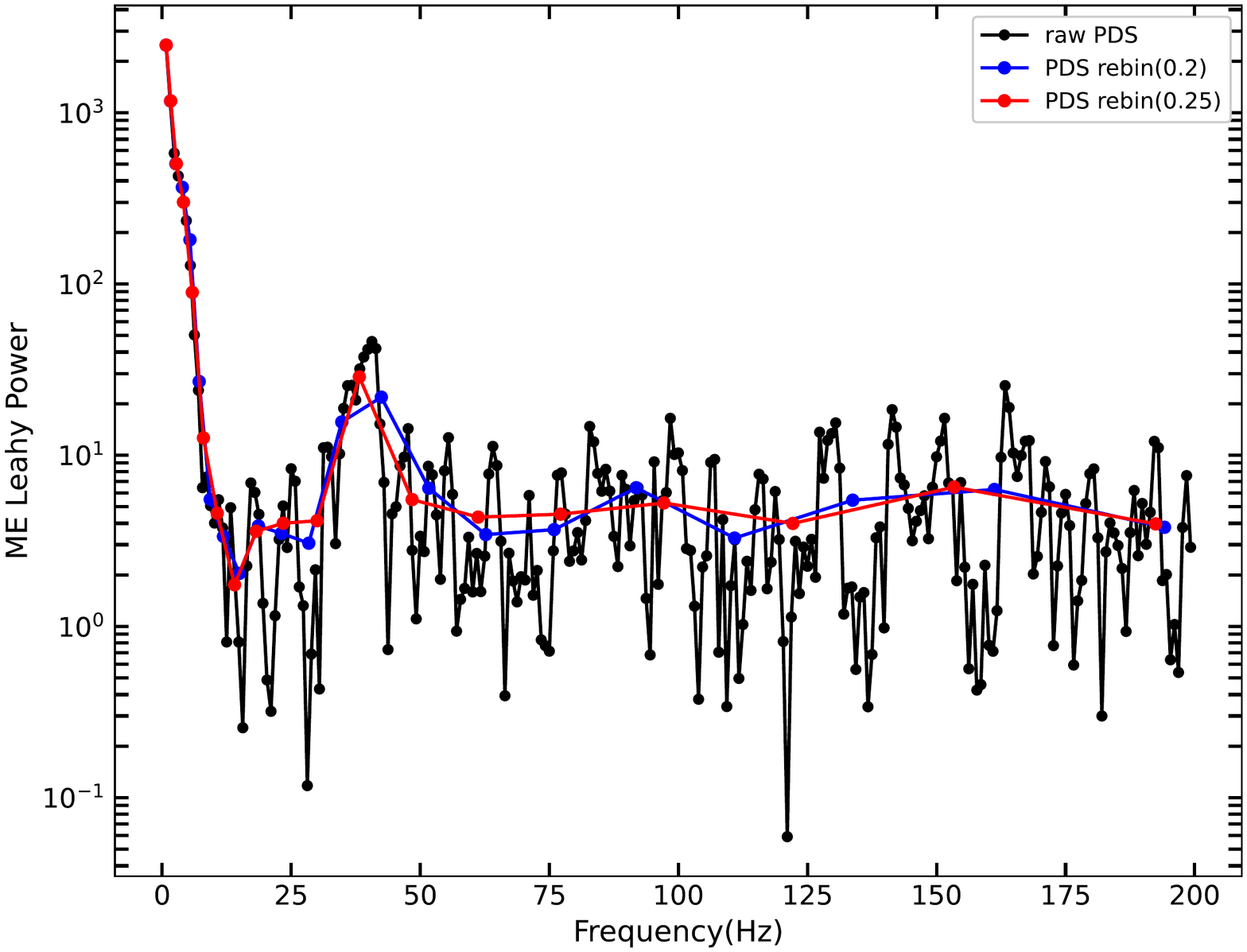}
\caption{
The normalized Leahy power of ME with the light curve of 1.28\,s. The black line is the raw PDS of ME, the blue and red lines are the logarithmic rebining of the raw PDS with factor of 0.2 and 0.25, respectively. The rebin factor of 0.25 is more suitable for the QPO signal.
\label{fig_mepdsrebin}}
\end{center}
\end{figure}

We use the same chosen noise model ($H_0$ model) to fit the periodogram and same samples of simulated periodograms in the previous step to search for the highest data/model outlier. We bin the periodogram geometrically, where the bin size grows with frequency and the bin factor is shown in Table \ref{tab:TR}. Figure \ref{fig_mepdsrebin} also presents as an example for the bin of 1.28\,s periodogram and we choose the bin factor as 0.25.
We also compare the observed value of the $T_{\rm{R}}$ statistic and compute the corresponding p-values for the three duration of the burst. The results are summarized in Table \ref{tab:TR}.
The QPO with central frequency at about 40\,Hz can be found for different duration. The logarithmic rebin of the periodogram can not ensure all the QPO frequency bins are in the same bin, therefore, this method are just used as a verification of the QPO.

\begin{table}
\footnotesize
\caption{Results for the significance of QPO and QPO parameters at different duration time using the $T_{\rm{R}}$ statistic.}
\scriptsize{}
\label{tab:TR}
\medskip
\begin{center}
\begin{tabular}{c c c c c c}
\hline \hline
Duration time(s) & $H_0$ model &  rebin factor & Frequency (Hz) &  $T_{\rm{R}}$ & p-value  \\
\hline
 0.32 & PL  & 0.2  & 37.3 & 10.6 & 0.0047 \\
\hline
 0.64 & BPL & 0.2 & 37.3 & 10.3 & 0.00096 \\
\hline
 1.28 & BPL & 0.25  & 38.2 & 9.8 & 0.00024 \\
\hline \hline
\end{tabular}
\end{center}
\end{table}

\subsection{Cross correlation analysis}

\begin{figure}
\begin{center}
\includegraphics[width=0.5\textwidth]{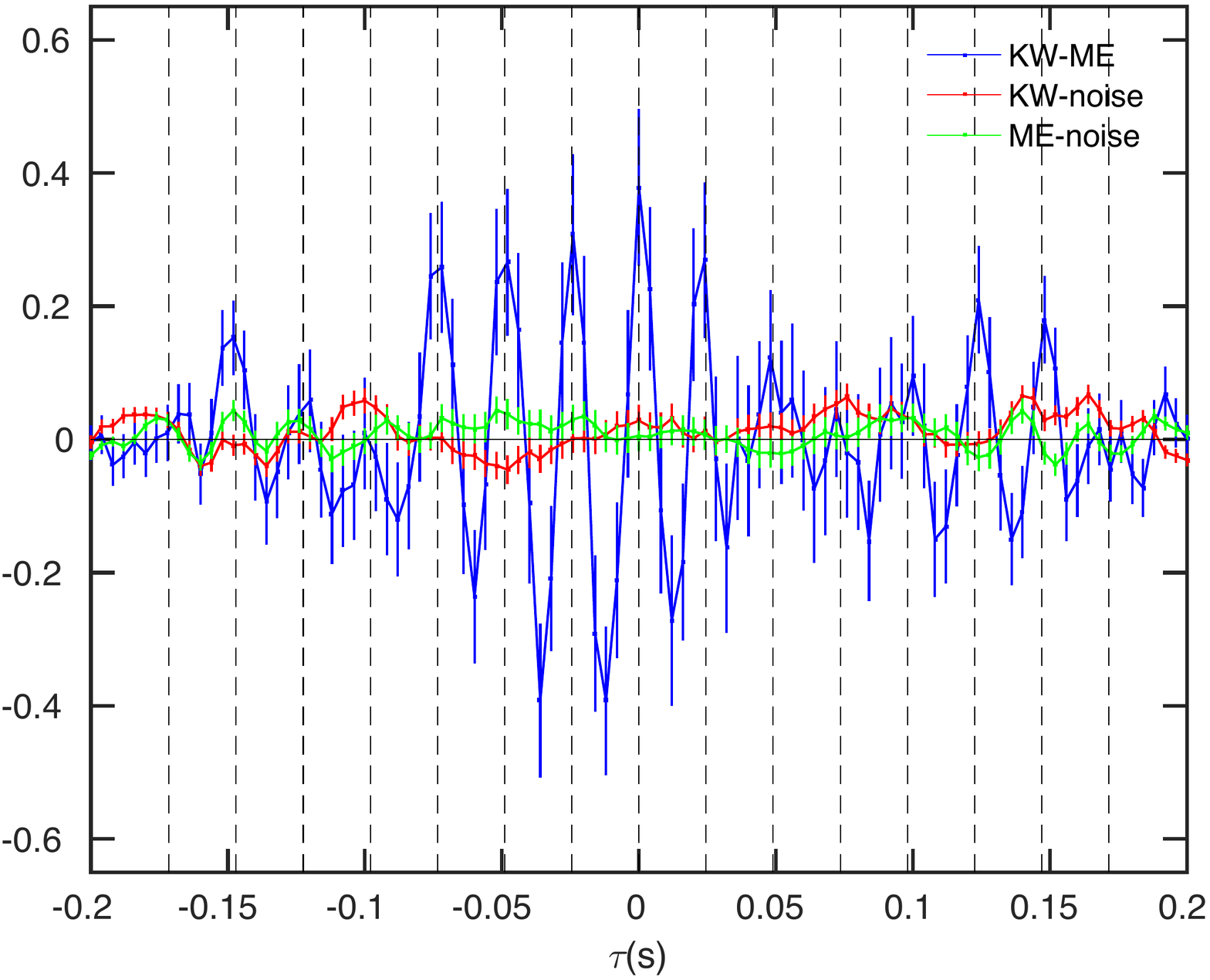}\caption{
Cross correlation between the light curves of {\sl Insight}-HXMT/ME (18--50\,keV) and KW (18--80\,keV). Blue points and solid line: coefficient of the cross correlation between the detrended light curves of {\sl Insight}-HXMT/ME and KW. Green points and line: between the detrended light curve of {\sl Insight}-HXMT/ME and white noise. Red points and line: between the detrended light curve of KW and white noise. The errors are estimated with Monte Carlo simulations. The vertical dashed lines mark the QPO period of $\sim$$24.5$\,ms.
\label{fig_KWME_corr}}
\end{center}
\end{figure}

To further verify the robustness of the $\sim$40\,Hz QPO detection, cross-correlations between the detrended light curves of \insight{}/ME and Konus-Wind are made and shown in Figure \ref{fig_KWME_corr}, along with cross-correlations between each light curve and white noise. The QPOs in both light curves match each other both in phase and period of 24.5\,ms, while no statistically meaningful correlation is found with white noise.

Cross correlation function (CCF) is used here to understand if the QPO signals found in the light curves of \insight{}/ME and Konus-Wind are synchronised. The CCF between the detrended light curves (after removing the three broad Gaussian peaks by fitting with eq. (\ref{eq0}) of ME and Konus-Wind shows strong oscillations of $\sim24.5$\,ms from $\tau=-0.1$\,s to $\tau=0.05$\,s as plotted in Figure \ref{fig_KWME_corr}. The errors for CCF are estimated from Monte-Carlo method, by sampling the light curves 1000 times. We also make the cross correlation between the detrended light curves and white noise with the same average counting rate, as expected, there are no oscillation structures in CCFs. However, there are some oscillation signals around $-0.15$\,s and $0.15$\,s, which are also generated by the oscillation structures in the real light curves.

\begin{figure}
\begin{center}
\includegraphics[width=0.5\textwidth]{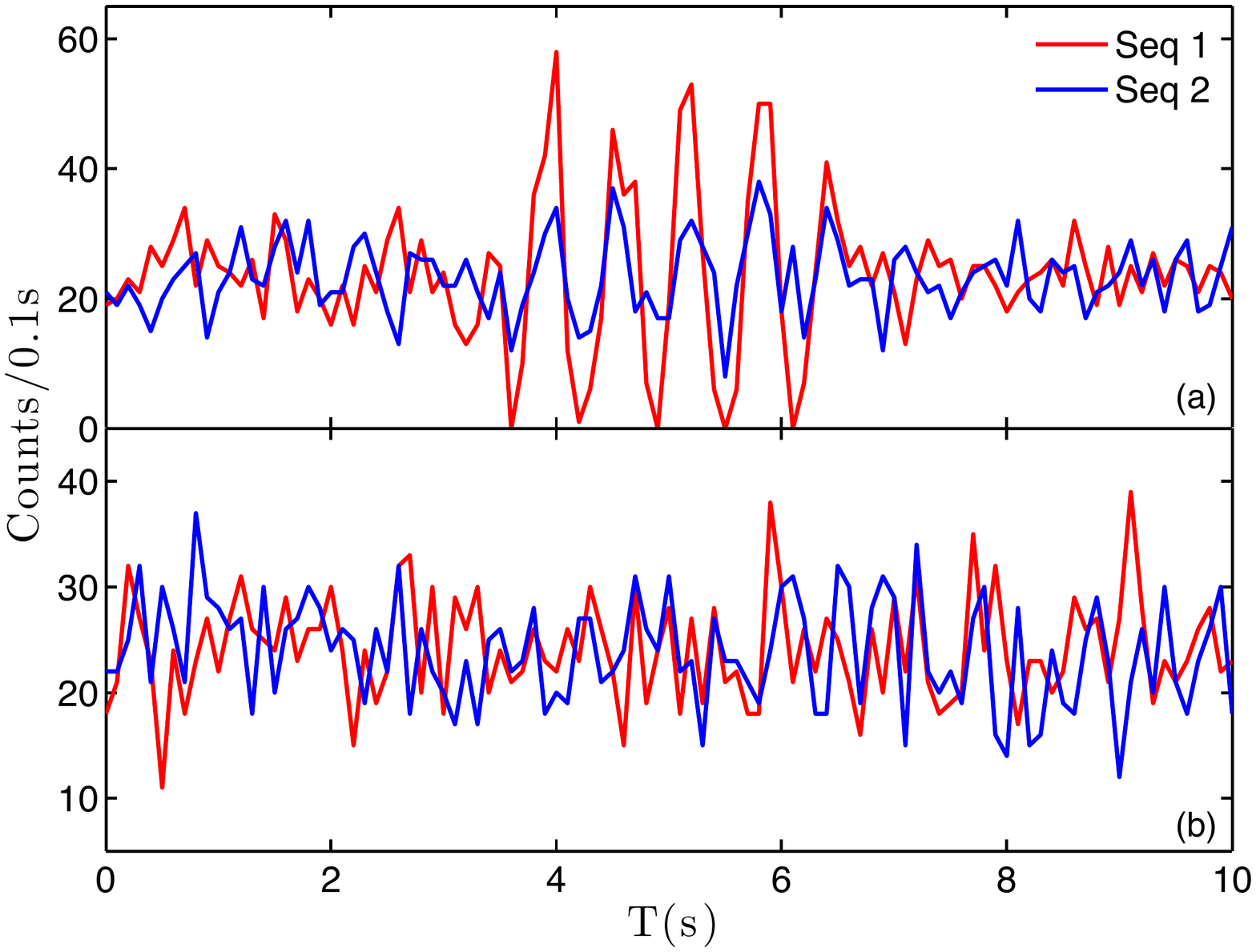}\caption{
Simulated light curves with sinusoidal signals and only white noise. {\bf Upper panel (a):} simulated light curves with 3 seconds of sinusoidal signals of amplitudes 8 and 24, respectively. {\bf Lower panel (b):} simulated light curves with only white noise with Poisson expectation value of 24 in each time bin.
\label{fig_sin_lc}}
\end{center}
\end{figure}

To further understand the above results of cross correlation analysis, we use sinusoidal signals to simulate the structures in CCFs obtained above. As shown in Figure \ref{fig_sin_lc} (upper panel: background + sinusoidal signals), two sinusoidal signals of amplitudes of 8 and 24 are simulated with background 24, which have frequency $10/2\pi$\,Hz and only appear for 3 seconds in the light curves. The Poisson fluctuations are added on the light curves. The background light curves only including white noise are generated from Poisson sampling with expected count of 24 in each time bin, as shown in Figure \ref{fig_sin_lc} (lower panel: background only) for two of the simulated background light curves of white noise.

\begin{figure}
\begin{center}
\includegraphics[width=0.5\textwidth]{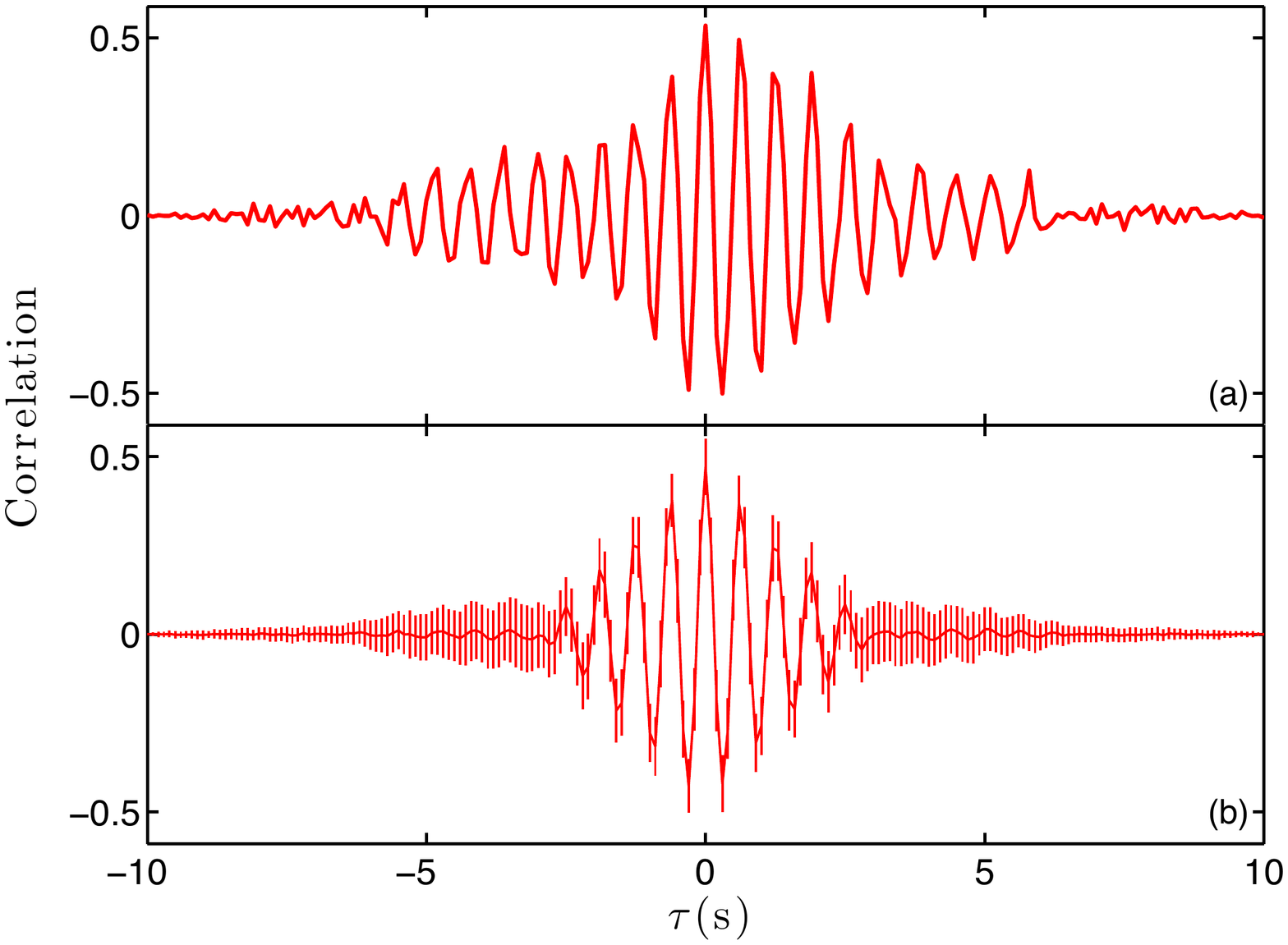}
\caption{
Cross correlations between the two simulated light curves with strong and weak sinusoidal oscillations. {\bf Upper panel (a):} One cross correlation for the two simulated light curves (blue and red lines in the upper panel of Figure \ref{fig_sin_lc}). {\bf Lower panel (b):} mean cross correlations for the 1000 simulated light curves. Therefore the strong sinusoidal oscillations are used as a template to identify effectively the weak sinusoidal oscillations, which can only be seen marginally alone.
\label{fig_sin_corr0}}
\end{center}
\end{figure}

We first make cross correlations between the two simulated light curves (including their backgrounds) with different amplitudes of sinusoidal oscillations. Very significant and periodic structures are shown in Figure \ref{fig_sin_corr0}, demonstrating the power of cross correlation analysis in identifying weakly oscillating signals (`Seq 2': blue line in the upper panel of Figure \ref{fig_sin_lc}), given a template of the oscillations (`Seq 1': red line in the upper panel of Figure \ref{fig_sin_lc}). Therefore, the cross correlation analysis presented in Figure \ref{fig_KWME_corr} is equivalent to using ME's light curve, which contains strong QPO signals, as a template in identifying the relatively weak QPOs in the light curve of Konus-Wind.

\begin{figure}
\begin{center}
\includegraphics[width=0.5\textwidth]{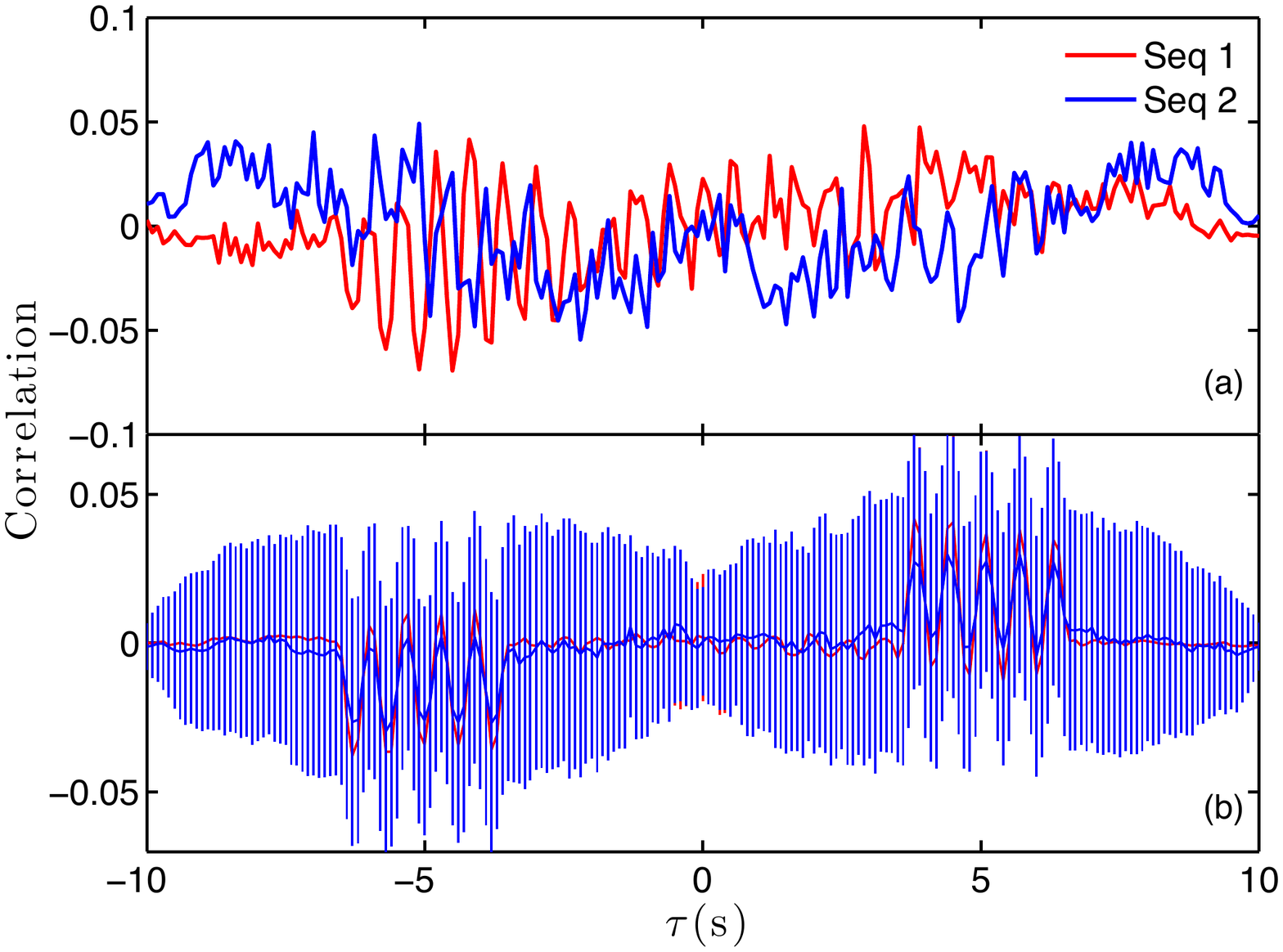}\caption{
The cross correlations for the simulated light curves without sinusoidal oscillations. \textbf{Upper panel (a):} One cross correlations for the simulated light curves. \textbf{Lower panel (b):} mean cross correlations for the 1000 simulated light curves.
\label{fig_sin_corr}}
\end{center}
\end{figure}

The CCFs between the white noise and both light curves with weak and strong oscillation signals are shown in Figure \ref{fig_sin_corr}. The upper panel shows the result from one single simulation run, where many oscillatory structures exist (Figure \ref{fig_sin_corr} upper panel). The processes are repeated for 1000 times to obtain the mean CCFs, which show clear oscillation structures around $\pm5$\ (lower panel in Figure \ref{fig_sin_corr}). These structures are similar and should have the same origin as the structures near both ends of the correlation functions of white noise with the light curves of ME and Konus-Wind, as shown in Figure \ref{fig_KWME_corr}.

\section{Physical interpretation }
\label{sec:physics}
\subsection{Constraints on theoretical models}

The 40\,Hz QPO reported in this paper poses the following constraints on the available models for FRB~200428:
\begin{itemize}
\item The fact that this burst is special with respect to the majority of SGR bursts suggest that the rarity of FRB-SGR burst associations \citep{2020arXiv200511479L} is likely due to intrinsic rather than extrinsic (beaming or narrow spectra; \citet{2020arXiv200511479L}) factors.


\item Since QPOs are directly related with crust oscillations, and since the two spikes on the X-ray lightcurve that are associated with the two pulses of FRB 200428 are roughly consistent with QPO peaks, it is almost certain that the X-ray emission as well as the associated FRB pulses originate from the magnetosphere of the neutron star \citep{lu2020,Wang20,2020arXiv200502100W,2020arXiv200616231W,2020arXiv200603468K,2020arXiv200603270Y}. Models invoking emission outside the magentosphere \citep{margalit2020,2020arXiv200600484Y} are thus disfavored, unless the ejections of the emitter are associated with crust activities \citep{2020arXiv200604649Y}.
\item The trigger of QPOs in the neutron star crust is likely from internal, but an external trigger powered by infall of an asteroid is also possible \citep{2020arXiv200512048D,2020arXiv200604601G}, which might explain the rarity of the FRB-associated bursts. Such models, however, need to interpret the existence of QPOs already before the spikes associated with FRB pulses.
\end{itemize}

\subsection{A plausible QPO model}

The identified QPO at 40 Hz may be explained as the toroidal mode of the crust \citep{Duncan98}, or the magnetoelastic modes if the coupling effect between the crust and core is considered \citep{Hoven2011}. For the crustal toroidal mode, the 40 Hz QPO can be explained as the $_3t_0$ mode if the tangled magnetic field in the crust is $B_t\ll B_{\mu}$,
or the fundamental toroidal mode $_2t_0$ if $B_t\approx B_{\mu}$,
where $_{\ell}t_n$ is the notation for the general toroidal modes, $\ell$ is the angular quantum number, and $n$ is the number of radial nodes in the eigenfunctions, and $B_{\mu}=(4\pi \mu)^{1/2}\approx 4\times 10^{15}\rho_{14}^{0.4}$ G with $\mu$ being the crust shear modulus, and $\rho_{14}$ being crust density in units of $10^{14}$g cm$^{-3}$ \citep{Duncan98}. Note the internal magnetic field is usually expected to be much higher than the surface dipole magnetic field, which is $2.2\times10^{14}$ G for SGR J1935+2154 \citep{israel2016}.

\section{Discussion and Summary}
\label{sec:summary}
Magnetar bursts are a potential window for the interior of neutron stars, via the oscillations measured in magnetar giant flares. As for the X-ray burst associated with FRB~200428, different strategies are used to search for the QPO and confirm the significance of the QPO.
Monte Carlo simulations of light curves fail to be predictive when there is no precise knowledge of the burst profile. Moreover, there is a degeneracy between the burst profile, a potential red noise component, and the QPO.

In the absence of the burst profile, we advocate a conservative Bayesian method which models the Leahy-normalized power of the burst as a pure red noise process.
The conventional method to generate Leahy-normalized power is used the whole light curves of the X-ray burst.
However, when the QPO is present in part of the burst phase, we may overestimate the significance of the QPO signal and underestimate the errors of QPO parameters.
The reliable way to estimate the significance of the QPO and its parameters in non-stationary processes is to use the time segment of data in which the QPO is present.

In the peculiar X-ray burst associated with FRB~200428 from the Galactic magnetar \sgrnos, we have discovered a QPO with a central frequency of $39.14_{-1.94}^{+1.88}$ \,Hz and width of $4.88_{-2.73}^{+5.67}$ between 18--50\,keV with p-value of 2.9e-4. The frequency, phase, modulation pattern and energy range all point to a physical connection with the two X-ray spikes coinciding with the two radio pulses of FRB~200428. The rarity of both QPOs in magnetar X-ray bursts and short radio pulses from magnetars indicates that the rarity of FRB-SGR burst associations \citep{2020arXiv200511479L} is very likely due to intrinsic reasons. It also poses important constraints on available FRB models proposed to interpret FRB~200428. In particular, models invoking relativistic shocks \citep{margalit2020} are disfavored. Instead, our QPO result supports these models invoking crust oscillation as the driver of magnetospheric emission from a magnetar magnetosphere \citep{lu2020,Wang20,2019ApJ...879....4W}. The specific frequency (40 Hz) of the QPOs may also pose interesting constraints on the magnetic field configurations in the neutron star crust.

\section{Acknowledgments}
 {\bf Acknowledgements} This work made use of the data from the \insight{} mission, a project funded by China National Space Administration (CNSA) and the Chinese Academy of Sciences (CAS).  We gratefully acknowledge the support from the National Program on Key Research and Development Project (Grant No.2021YFA0718500) from the Minister of Science and Technology of China (MOST). The authors thank supports from the National Natural Science Foundation of China under Grants U1838201, U1838202, U1938109, U1938102, U1938108, 11473027, 11733009, 1173309, Y829113, 11673023, 11703002, 11833003. We also thank the support from the Strategic Priority Program on Space Science, the Chinese Academy of Sciences, Grant No.XDA15020503.
 J.S.Wang acknowledges the support from the Alexander von Humboldt Foundation.

\bibliography{mainref}

\end{document}